\documentclass[11pt,a4paper]{article}
\pdfoutput=1 % if your are submitting a pdflatex (i.e. if you have
\RequirePackage{ifpdf} 
\usepackage{amsmath} 
\usepackage{mathtools}

\usepackage{jheppub}
\usepackage{pstricks}
\usepackage[final]{pdfpages} 
\usepackage{ifpdf} 
\usepackage{slashed}

\usepackage{hyperref}

\usepackage{color} 
\usepackage{graphics}

\usepackage{etoolbox} % Load before 'breqn' package to make 'lineno' work
\usepackage{fixmath}

\usepackage{notoccite} % If you have \cite commands in \section-like commands, or in \caption, the
\usepackage{caption} 
\usepackage{subcaption} 
\usepackage{amsfonts}

\usepackage[normalem]{ulem}

\title{Pseudo-scalar Form Factors at Three Loops in QCD}

\author[a]{Taushif Ahmed,} \author[b]{Thomas Gehrmann,}
\author[c]{Prakash Mathews,} \author[a]{Narayan Rana} \author[a]{and\\
  V. Ravindran}

\affiliation[a]{The Institute of Mathematical Sciences, IV Cross Road,
  CIT Campus, \\Chennai 600 113, Tamil Nadu, India}
\affiliation[b]{Department of Physics, University of Z\"urich,
  Winterthurerstrasse 190, \\CH-8057 Z\"urich, Switzerland}
\affiliation[c]{Saha Institute of Nuclear Physics, 1/AF Bidhan Nagar,
  Kolkata 700 064, \\West Bengal, India}

\emailAdd{taushif@imsc.res.in} \emailAdd{thomas.gehrmann@uzh.ch}
\emailAdd{prakash.mathews@saha.ac.in} \emailAdd{rana@imsc.res.in}
\emailAdd{ravindra@imsc.res.in}

\abstract{ The coupling of a pseudo-scalar Higgs boson to gluons is
  mediated through a heavy quark loop. In the limit of large quark mass,
  it is described by an effective Lagrangian that only admits light
  degrees of freedom. In this effective theory, we compute the
  three-loop massless QCD corrections to the form factor that
  describes the coupling of a pseudo-scalar Higgs boson to gluons. Due
  to the axial anomaly, the pseudo-scalar operator for the gluonic
  field strength mixes with the divergence of the axial vector
  current.  Working in dimensional regularization and using the
  't~Hooft-Veltman prescription for the axial vector current, we
  compute the three-loop pseudo-scalar form factors for massless
  quarks and gluons.  Using the universal infrared factorization
  properties, we independently derive the three-loop operator mixing
  and finite operator renormalisation from the renormalisation group
  equation for the form factors, thereby confirming recent results in
  the operator product expansion. The finite part of the three-loop
  form factor is an important ingredient to the precise prediction of
  the pseudo-scalar Higgs boson production cross section at hadron
  colliders. We discuss potential applications and derive the hard
  matching coefficient in soft-collinear effective theory. }
\preprint{} 

\begin{document}
\allowdisplaybreaks[4]
\unitlength1cm
\maketitle
\flushbottom

%************
% Definition
%************

\def\D{{\cal D}}
\def\DD{\overline{\cal D}}
\def\g{\overline{\cal G}}
\def\gm{\gamma}
\def\M{{\cal M}}
\def\ep{\epsilon}
\def\epm1{\frac{1}{\epsilon}}
\def\epm2{\frac{1}{\epsilon^{2}}}
\def\epm3{\frac{1}{\epsilon^{3}}}
\def\epm4{\frac{1}{\epsilon^{4}}}
\def\unM{\hat{\cal M}}
\def\ashat{\hat{a}_{s}}
\def\asmur{a_{s}^{2}(\mu_{R}^{2})}
\def\sigbar{{{\overline {\sigma}}}\left(a_{s}(\mu_{R}^{2}), L\left(\mu_{R}^{2}, m_{H}^{2}\right)\right)}
\def\sigbarn{{{{\overline \sigma}}_{n}\left(a_{s}(\mu_{R}^{2}) L\left(\mu_{R}^{2}, m_{H}^{2}\right)\right)}}
\def\unas{ \left( \frac{\hat{a}_s}{\mu_0^{\epsilon}} S_{\epsilon} \right) }
\def\rnM{{\cal M}}
\def\bt{\beta}
\def\cD{{\cal D}}
\def\cC{{\cal C}}
\def\ca{\text{\tiny C}_\text{\tiny A}}
\def\cf{\text{\tiny C}_\text{\tiny F}}
\def\ct{{\red []}}
\def\sv{\text{SV}}
\def\murOmu{\left( \frac{\mu_{R}^{2}}{\mu^{2}} \right)}
\def\bb{b{\bar{b}}}
\def\bt0{\beta_{0}}
\def\bt1{\beta_{1}}
\def\bt2{\beta_{2}}
\def\bt3{\beta_{3}}
\def\gm0{\gamma_{0}}
\def\gm1{\gamma_{1}}
\def\gm2{\gamma_{2}}
\def\gm3{\gamma_{3}}
\def\nn{\nonumber}
\def\l{\left}
\def\r{\right}

\newcommand{\dis}{}
\newcommand{\overbar}[1]{mkern-1.5mu\overline{\mkern-1.5mu#1\mkern-1.5mu}\mkern
1.5mu}

%*******
% Intro
%*******

\section{Introduction}
\setcounter{equation}{0}
\label{sec:intro}

Form factors are the matrix elements of local composite operators
between physical states. In the calculation of scattering cross
sections, they provide the purely virtual corrections.  For example,
in the context of hard scattering processes such as
Drell-Yan~\cite{Altarelli:1979ub,Hamberg:1990np} and the Higgs boson
production in gluon
fusion~\cite{Dawson:1990zj,Djouadi:1991tka,Graudenz:1992pv,Spira:1995rr,
  Djouadi:1995gt, Spira:1997dg,Catani:2001ic, Harlander:2002wh,
  Anastasiou:2002yz, Ravindran:2003um, Ravindran:2004mb,
  Harlander:2005rq,Anastasiou:2015ema}, the form factors corresponding
to the vector current operator $\overline \psi \gamma_\mu \psi$ and
the gluonic operator $G^{a}_{\mu \nu} G^{a,\mu\nu}$ contribute,
respectively.  Here $\psi$ is the fermionic field operator and
$G^{a}_{\mu \nu}$ is the field tensor of the non-Abelian gauge field
$A_\mu^a$ corresponding to the gauge group SU(N).  In Quantum
Chromodynamics (QCD) the form factors can be computed order by order
in the strong coupling constant using perturbation theory.  Beyond
leading order, the ultraviolet (UV) renormalisation of the form
factors involves the renormalisation of the composite operator itself,
besides the standard procedure for coupling constant and external
fields.

The resulting UV finite form factors still contain divergences of
infrared (IR) origin, namely, soft and collinear divergences due to
the presence of massless gluons and quarks/antiquarks in the theory.
The inclusive hard scattering cross sections require, in addition to
the form factor, the real-emission partonic subprocesses as well as
suitable mass factorisation kernels for incoming partons.  The soft
divergences in the form factor resulting from the gluons cancel
against those present in the real emission processes and the mass
factorisation kernels remove the remaining collinear divergences
rendering the hadronic inclusive cross section IR finite.  While these
IR divergences cancel among various parts in the perturbative
computations, they can give rise to logarithms involving physical
scales and kinematic scaling variables of the processes under study.
In kinematical regions where these logarithms become large, they may
affect the convergence and reliability of the perturbation series
expansion in powers of the coupling constant.  The solution for this
problem goes back to the pioneering work by
Sudakov~\cite{Sudakov:1954sw} on the asymptotic behaviour of the form
factor in Quantum Electrodynamics: all leading logarithms can be
summed up to all orders in perturbation theory.  Later on, this
resummation was extended to non-leading
logarithms~\cite{Collins:1980ih} and systematised for non-Abelian
gauge theories~\cite{Sen:1981sd}.  Ever since, form factors have been
central to understand the underlying structure of amplitudes in gauge
theories.

The infrared origin of universal logarithmic corrections to form
factors~\cite{Magnea:1990zb} and scattering amplitudes results in a
close interplay between resummation and infrared pole structure.
Working in dimensional regularisation in $d=4+\epsilon$ dimensions,
these poles appear as inverse powers in the Laurent expansion in
$\epsilon$. In a seminal paper, Catani~\cite{Catani:1998bh} proposed a
universal formula for the IR pole structure of massless two-loop QCD
amplitudes of arbitrary multiplicity (valid through to double pole
terms).  This formula was later on justified systematically from
infrared factorization~\cite{Sterman:2002qn}, thereby also revealing
the structure of the single poles in terms of the anomalous dimensions
for the soft radiation.  In \cite{Ravindran:2004mb}, it was shown that
the single pole term in quark and gluon form factors up to two loop
level can be shown to decompose into UV ($\gamma_{I}, I=q,g$) and
universal collinear ($B_I$), color singlet soft ($f_I$) anomalous
dimensions, later on observed to hold even at three loop level
in~\cite{Moch:2005tm}.  An all loop conjecture for the pole structure
of the on-shell multi-loop multi-leg amplitudes in SU(N) gauge theory
with $n_f$ light flavors in terms of cusp ($A_I$), collinear ($B_{I}$)
and soft anomalous dimensions ($\Gamma_{IJ},f_I$ - colour non-singlet
as well as singlet) was proposed by Becher and
Neubert~\cite{Becher:2009cu} and Gardi and Magnea~\cite{Gardi:2009qi},
generalising the earlier results~\cite{Catani:1998bh,Sterman:2002qn}.
The validity of this conjecture beyond three loops depends on the
presence/absence of non-trivial colour correlations and crossing
ratios involving kinematical invariants~\cite{Almelid:2015jia} and
there exists no all-order proof at present. The three-loop expressions
for cusp, collinear and colour singlet soft anomalous dimensions were
extracted~\cite{Moch:2005ba,Laenen:2005uz} from the three loop flavour
singlet~\cite{Vogt:2004mw} and non-singlet~\cite{Moch:2004pa}
splitting functions, thereby also predicting~\cite{Moch:2005tm} the
full pole structure of the three-loop form factors.

The three-loop quark and gluon form factors through to finite terms
were computed
in~\cite{Baikov:2009bg,Gehrmann:2010ue,Gehrmann:2011xn,Gehrmann:2014vha}
and subsequently extended to higher powers in the $\epsilon$
expansion~\cite{Gehrmann:2010tu}.  These results were enabled by
modern techniques for multi-loop calculations in quantum field theory,
in particular integral reduction methods.  These are based on
integration-by-parts (IBP)~\cite{Tkachov:1981wb,Chetyrkin:1981qh} and
Lorentz invariance (LI)~\cite{Gehrmann:1999as} identities which reduce
the set of thousands of multi-loop integrals to the one with few
integrals, so called master integrals (MIs) in dimensional
regularisation. To solve these large systems of IBP and LI identities,
the Laporta algorithm~\cite{Laporta:2001dd}, which is based on
lexicographic ordering of the integrals, is the main tool of
choice. It has been implemented in several computer algebra
codes~\cite{Anastasiou:2004vj, Smirnov:2008iw, Studerus:2009ye,
  vonManteuffel:2012np, Lee:2012cn,Lee:2013mka}.  The MIs relevant to
the form factors are single-scale three-loop vertex functions, for
which analytical expressions were derived in
Refs.~\cite{Gehrmann:2005pd, Gehrmann:2006wg, Heinrich:2007at,
  Heinrich:2009be, Lee:2010cga, Gehrmann:2010ue}.
  
Recently, some of us have applied these state-of-the-art methods to
accomplish the task of computing spin-2 quark and gluon form factors
up to three loops~\cite{Ahmed:2015qia} level in SU(N) gauge theory
with $n_f$ light flavours.  These form factors are ingredients to the
precise description of production cross sections for graviton
production, that are predicted in extensions of the Standard Model.
In addition, the spin-2 form factors relate to operators with higher
tensorial structure and thus provide the opportunity to test the
versatility and robustness of calculational techniques for the vertex
functions at three loop level.  The results~\cite{Ahmed:2015qia} also
confirm the universality of the UV and IR structure of the gauge
theory amplitudes in dimensional regularisation.

In the present work, we derive the three-loop corrections to the quark
and gluon form factors for pseudo-scalar operators. These operators
appear frequently in effective field theory descriptions of extensions
of the Standard Model. Most notably, a pseudo-scalar state coupling to
massive fermions is an inherent prediction of any two-Higgs doublet
model~\cite{Fayet:1974pd,Fayet:1976et, Fayet:1977yc,
  Dimopoulos:1981zb, Sakai:1981gr,Inoue:1982pi, Inoue:1983pp,
  Inoue:1982ej}.  In the limit of infinite fermion mass, this gives
rise to the operator insertions considered here.  The recent discovery
of a Standard-Model-like Higgs boson at the LHC~\cite{Aad:2012tfa,
  Chatrchyan:2012xdj} has not only revived the interest in such Higgs
bosons but also prompted the study of the properties of the discovered
boson to identify either with lightest scalar or pseudo-scalar Higgs
bosons of extended models. Such a study requires precise predictions
for their production cross sections. In the context of a CP-even
scalar Higgs boson, results for the inclusive production cross section
in the gluon fusion are available up to N$^{3}$LO
QCD~\cite{Anastasiou:2002yz,Harlander:2002wh,Ravindran:2003um,
  Anastasiou:2015ema}, based on an effective scalar coupling that
results from integration of massive quark loops that mediate the
coupling of the Higgs boson to
gluons~\cite{Ellis:1975ap,Shifman:1979eb,Kniehl:1995tn}.  On the other
hand for the CP-odd pseudo scalar, only NNLO QCD
results~\cite{Kauffman:1993nv,Djouadi:1993ji,Harlander:2002vv,Anastasiou:2002wq,
  Ravindran:2003um} in the effective theory~\cite{Chetyrkin:1998mw}
are known.  The exact quark mass dependence for scalar and
pseudo-scalar production is known to NLO
QCD~\cite{Spira:1993bb,Spira:1995rr}, and is usually included through
a re-weighting of the effective theory results. Soft gluon resummation
of the gluon fusion cross section has been performed to N$^3$LL for
the scalar
case~\cite{Catani:2003zt,Moch:2005ky,Ravindran:2005vv,Ravindran:2006cg,
  Idilbi:2005ni,Ahrens:2008nc,deFlorian:2009hc,Bonvini:2014joa,Catani:2014uta}
and to NNLL for the pseudo-scalar case~\cite{deFlorian:2007sr}. A
generic threshold resummation formula valid to N$^3$LL accuracy for
colour-neutral final states was derived in~\cite{Catani:2014uta},
requiring only the virtual three-loop amplitudes as process-dependent
input.  The numerical impact of soft gluon resummation in scalar and
pseudo-scalar Higgs boson production and its combination with mass
corrections is reviewed comprehensively in~\cite{Schmidt:2015cea}.
The three-loop corrections to the pseudo-scalar form factors computed
in this article are an important ingredient to the N$^3$LO and N$^3$LL
gluon fusion cross sections~\cite{Ahmed:2015pSSV} for pseudo-scalar
Higgs boson production, thereby enabling predictions at the same level
of precision that is attained in the scalar case.
   
The framework of the calculation is outlined in
Section~\ref{sec:frame}, where we describe the effective
theory~\cite{Chetyrkin:1998mw}.  Due to the pseudo-scalar coupling,
one is left with two effective operators with same quantum number and
mass dimensions, which mix under renormalisation.  Since these
operators contain the Levi-Civita tensor as well as $\gamma_5$, the
computation of the matrix elements requires additional care in
$4+\epsilon$ dimensions where neither Levi-Civita tensor nor
$\gamma_5$ can be defined unambiguously.  We use the prescription by
't Hooft and Veltman~\cite{'tHooft:1972fi,Larin:1993tq} to define
$\gamma_5$.  We describe the calculation in Section~\ref{sec:FF},
putting particular emphasis on the UV renormalisation.  Exploiting the
universal IR pole structure of the form factors, we determine the UV
renormalisation constants and mixing of the effective operators up to
three loop level.  We also show that the finite renormalisation
constant, known up to three loops~\cite{Larin:1993tq}, required to
preserve one loop nature of the chiral anomaly, is consistent with
anomalous dimensions of the overall renormalisation constants. As a
first application of our form factors, we compute the hard matching
functions for N$^3$LL resummation in soft-collinear effective theory
(SCET) in Section~\ref{sec:scet}.  Section~\ref{sec:conc} summarises
our results and contains an outlook on future applications to
precision phenomenology of pseudo-scalar Higgs production.

% ******* Theory *******

\section{Framework of the Calculation}
\label{sec:frame}
\subsection{The Effective Lagrangian}
\label{sec:ThreResu}

A pseudo-scalar Higgs boson couples to gluons only indirectly through
a virtual heavy quark loop. This loop can be integrated out in the
limit of infinite quark mass.  The resulting effective
Lagrangian~\cite{Chetyrkin:1998mw} encapsulates the interaction
between a pseudo-scalar $\Phi^A$ and QCD particles and reads:
\begin{align} {\cal L}^{A}_{\rm eff} = \Phi^{A}(x) \Big[ -\frac{1}{8}
  {C}_{G} O_{G}(x) - \frac{1}{2} {C}_{J} O_{J}(x)\Big]
\end{align}
where the operators are defined as
\begin{equation}
  O_{G}(x) = G^{\mu\nu}_a \tilde{G}_{a,\mu
    \nu} \equiv  \epsilon_{\mu \nu \rho \sigma} G^{\mu\nu}_a G^{\rho
    \sigma}_a\, ,\qquad
  O_{J}(x) = \partial_{\mu} \left( \bar{\psi}
    \gamma^{\mu}\gamma_5 \psi \right)  \,.
  \label{eq:operators}
\end{equation}
The Wilson coefficients $C_G$ and $C_J$ are obtained by integrating
out the heavy quark loop, and $C_G$ does not receive any QCD
corrections beyond one loop due to the Adler-Bardeen
theorem~\cite{Adler:1969gk}, while $C_J$ starts only at second order
in the strong coupling constant. Expanded in
$a_s \equiv {g}_{s}^{2}/(16\pi^{2}) = \alpha_s/(4\pi)$,
they read
\begin{align}
  \label{eq:const}
  & C_{G} = -a_{s} 2^{\frac{5}{4}} G_{F}^{\frac{1}{2}} {\rm \cot} \beta
    \nonumber\\
  & C_{J} = - \left[ a_{s} C_{F} \left( \frac{3}{2} - 3\ln
    \frac{\mu_{R}^{2}}{m_{t}^{2}} \right) + a_s^2 C_J^{(2)} + \cdots \right] C_{G}\, .
\end{align}
In the above expressions, $G^{\mu\nu}_{a}$ and $\psi$ represent
gluonic field strength tensor and light quark fields, respectively and
$G_{F}$ is the Fermi constant and ${\rm \cot}\beta$ is the mixing angle in a generic
Two-Higgs-Doublet model.
$a_{s} \equiv a_{s} \left( \mu_{R}^{2} \right)$ is the strong coupling
constant renormalised at the scale $\mu_{R}$ which is related to the
unrenormalised one, ${\hat a}_{s} \equiv {\hat g}_{s}^{2}/(16\pi^{2})$
through

\begin{align}
  \label{eq:asAasc}
  {\hat a}_{s} S_{\epsilon} = \left( \frac{\mu^{2}}{\mu_{R}^{2}}  \right)^{\epsilon/2}
  Z_{a_{s}} a_{s}
\end{align}
with
$S_{\epsilon} = {\rm exp} \left[ (\gamma_{E} - \ln 4\pi)\epsilon/2
\right]$
and $\mu$ is the scale introduced to keep the strong coupling constant
dimensionless in $d=4+\epsilon$ space-time dimensions.  The
renormalisation constant $Z_{a_{s}}$~\cite{Tarasov:1980au} is given by
\begin{align}
  \label{eq:Zas}
  Z_{a_{s}}&= 1+ a_s\left[\frac{2}{\epsilon} \beta_0\right]
             + a_s^2 \left[\frac{4}{\epsilon^2 } \beta_0^2
             + \frac{1}{\epsilon}  \beta_1 \right]
             + a_s^3 \left[\frac{8}{ \epsilon^3} \beta_0^3
             +\frac{14}{3 \epsilon^2}  \beta_0 \beta_1 +  \frac{2}{3
             \epsilon}   \beta_2 \right]
\end{align}
up to ${\cal O}(a_{s}^{3})$. $\beta_{i}$ are the coefficients of the
QCD $\beta$ functions which are given by~\cite{Tarasov:1980au}
\begin{align}
  \beta_0&={11 \over 3 } C_A - {4 \over 3 } n_f T_{F}\, ,
           \nonumber \\[0.5ex]
  \beta_1&={34 \over 3 } C_A^2- 4 n_f C_F T_{F} -{20 \over 3} n_f
           T_{F} C_A \, ,
           \nonumber \\[0.5ex]
  \beta_2&={2857 \over 54} C_A^3 
           -{1415 \over 27} C_A^2 n_f T_{F}
           +{158\over 27} C_A n_f^2 T_{F}^{2}
           % \nonumber\\[0.5ex]
           % &
               +{44 \over 9} C_F n_f^2 T_{F}^{2}
               \nonumber\\
&-{205 \over 9} C_F C_A n_f T_{F}
               + 2 C_F^2 n_f T_{F}
\end{align}
with the SU(N) QCD color factors
\begin{equation}
  C_A=N,\quad \quad C_F={N^2-1 \over 2 N} \quad \text{and} \quad T_{F}= \frac{1}{2}\,.
\end{equation}
$n_f$ is the number of active light quark flavors.

\subsection{Treatment of $\gamma_5$ in Dimensional Regularization}
\label{sec:gamma5}

Higher order calculations of chiral quantities in dimensional regularization face the problem of
defining a generalization of
the inherently four-dimensional objects  $\gamma_5$ and $\varepsilon^{\mu\nu\rho\sigma}$
to values of $d\neq 4$. 
In this article, we have followed the most practical and
self-consistent definition of $\gamma_{5}$ for multiloop calculations
in dimensional regularization which was introduced by 't~Hooft and
Veltman through \cite{'tHooft:1972fi}
\begin{align}
  \gamma_5 = i \frac{1}{4!} \varepsilon_{\nu_1 \nu_2 \nu_3 \nu_4}
  \gamma^{\nu_1}  \gamma^{\nu_2} \gamma^{\nu_3} \gamma^{\nu_4} \,.
\end{align}
Here, $\varepsilon^{\mu\nu\rho\sigma}$ is the Levi-Civita tensor which is contracted as
\begin{align}
  \label{eqn:LeviContract}
  \varepsilon_{\mu_1\nu_1\lambda_1\sigma_1}\,\varepsilon^{\mu_2\nu_2\lambda_2\sigma_2}=
  \large{\left |
  \begin{array}{cccc}
    \delta_{\mu_1}^{\mu_2} &\delta_{\mu_1}^{\nu_2}&\delta_{\mu_1}^{\lambda_2} & \delta_{\mu_1}^{\sigma_2}\\
    \delta_{\nu_1}^{\mu_2}&\delta_{\nu_1}^{\nu_2}&\delta_{\nu_1}^{\lambda_2}&\delta_{\nu_1}^{\sigma_2}\\
    \delta_{\lambda_1}^{\mu_2}&\delta_{\lambda_1}^{\nu_2}&\delta_{\lambda_1}^{\lambda_2}&\delta_{\lambda_1}^{\sigma_2}\\
    \delta_{\sigma_1}^{\mu_2}&\delta_{\sigma_1}^{\nu_2}&\delta_{\sigma_1}^{\lambda_2}&\delta_{\sigma_1}^{\sigma_2}
  \end{array}
                                                                                       \right |}
\end{align}
and all the Lorentz indices are considered to be $d$-dimensional~\cite{Larin:1993tq}. 
In this scheme, a finite renormalisation of the 
axial vector current is required in order to fulfill chiral Ward identities and the Adler-Bardeen theorem. We discuss this 
in detail in Section~\ref{ss:UV} below. 

\section{Pseudo-scalar Quark and Gluon Form Factors}
\label{sec:FF}

The quark and gluon form factors describe the QCD loop corrections to
the transition matrix element from a color-neutral operator $O$ to an
on-shell quark-antiquark pair or to two gluons.  For the pseudo-scalar
interaction, we need to consider the two operators $O_{G}$ and
$O_{J}$, defined in Eq.~(\ref{eq:operators}), thus yielding in total
four form factors.
We define the unrenormalised gluon form factors at
${\cal O}({\hat a}_{s}^{n})$ as

\begin{align}
  \label{eq:DefFg}
  {\hat{\cal F}}^{G,(n)}_{g} \equiv \frac{\langle{\hat{\cal
  M}}^{G,(0)}_{g}|{\hat{\cal M}}^{G,(n)}_{g}\rangle}{\langle{\hat{\cal
  M}}^{G,(0)}_{g}|{\hat{\cal M}}^{G,(0)}_{g}\rangle}\, ,
  \qquad \qquad
  {\hat{\cal F}}^{J,(n)}_{g} \equiv \frac{\langle{\hat{\cal
  M}}^{G,(0)}_{g}|{\hat{\cal M}}^{J,(n+1)}_{g}\rangle}{\langle{\hat{\cal
  M}}^{G,(0)}_{g}|{\hat{\cal M}}^{J,(1)}_{g}\rangle}
\end{align}
and similarly the unrenormalised quark form factors through
\begin{align}
  \label{eq:DefFq}
  {\hat{\cal F}}^{G,(n)}_{q} \equiv \frac{\langle{\hat{\cal
  M}}^{J,(0)}_{q}|{\hat{\cal M}}^{G,(n+1)}_{q}\rangle}{\langle{\hat{\cal
  M}}^{J,(0)}_{q}|{\hat{\cal M}}^{G,(1)}_{q}\rangle}\, ,
  \qquad \qquad
  {\hat{\cal F}}^{J,(n)}_{q} \equiv \frac{\langle{\hat{\cal
  M}}^{J,(0)}_{q}|{\hat{\cal M}}^{J,(n)}_{q}\rangle}{\langle{\hat{\cal
  M}}^{J,(0)}_{q}|{\hat{\cal M}}^{J,(0)}_{q}\rangle}
\end{align}
where, $n=0, 1, 2, 3, \ldots$\,. In the above expressions
$|{\hat{\cal M}}^{\lambda,(n)}_{\beta}\rangle$ is the
${\cal O}({\hat a}_{s}^{n})$ contribution to the unrenormalised matrix
element for the transition from the bare operator $[O_{\lambda}]_B$
$(\lambda = G,J)$ to a quark-antiquark pair ($\beta=q$) or to two
gluons ($\beta=g$).  The expansion of these quantities in powers of
${\hat a}_{s}$ is performed through

\begin{tabular}{p{5cm}p{8.5cm}}
  \begin{align*}
    |{\cal M}^{\lambda}_{\beta}\rangle \equiv \sum_{n=0}^{\infty} {\hat
    a}^{n}_{s} S^{n}_{\epsilon}
    |{\hat{\cal M}}^{\lambda,(n)}_{\beta} \rangle
  \end{align*}
  &
    \begin{equation}
      \label{eq:Mexp}
      \hspace{-0.5cm}\text{and}\qquad
      {\cal F}^{\lambda}_{\beta} \equiv
      \sum_{n=0}^{\infty} \left[ {\hat a}_{s}^{n}
        \left( \frac{Q^{2}}{\mu^{2}} \right)^{n\frac{\epsilon}{2}}
        S_{\epsilon}^{n}  {\hat{\cal F}}^{\lambda,(n)}_{\beta}\right]\,.
    \end{equation}
\end{tabular}
\\
where, $Q^{2}=-2\, p_{1}.p_{2}$ and $p_{i}'s$ $(p_{i}^{2}=0)$ are the
momenta of the external quarks and gluons. Note that
$|{\hat{\cal M}}^{G,(n)}_{q}\rangle$ and
$|{\hat{\cal M}}^{J,(n)}_{g}\rangle$ start from $n=1$ i.e. from one
loop level.

\subsection{Calculation of the Unrenormalised Form Factors}
\label{ss:CalcFF}

The calculation of the unrenormalised pseudo-scalar form factors up to
three loops follows closely the steps used in the derivation of the
three-loop scalar and vector form factors
\cite{Gehrmann:2010ue,Gehrmann:2014vha}.  The Feynman diagrams for all
transition matrix elements (Eq.~(\ref{eq:DefFg}),
Eq.~(\ref{eq:DefFq})) are generated using
QGRAF~\cite{Nogueira:1991ex}. The numbers of diagrams contributing to
three loop amplitudes are 1586 for
$|{\hat{\cal M}}^{G,(3)}_{g}\rangle$, 447 for
$|{\hat{\cal M}}^{J,(3)}_{g}\rangle$, 400 for
$|{\hat{\cal M}}^{G,(3)}_{q}\rangle$ and 244 for
$|{\hat{\cal M}}^{J,(3)}_{q}\rangle$ where all the external particles
are considered to be on-shell. The raw output of QGRAF is converted to
a format suitable for further manipulation. A set of in-house routines
written in the symbolic manipulating program FORM
\cite{Vermaseren:2000nd} is utilized to perform the simplification of
the matrix elements involving Lorentz and color indices. Contributions
arising from ghost loops are taken into account as well since we use
Feynman gauge for internal gluons. For the external on-shell gluons,
we ensure the summing over only transverse polarization states by
employing an axial polarization sum:
\begin{equation}
  \label{eq:PolSum}
  \sum_{s} {\varepsilon^{\mu}}^{\, *}(p_{i},s)
  \varepsilon^{\nu}(p_{i},s)  = - \eta^{\mu\nu} + \frac{p_{i}^{\mu}
    q_{i}^{\nu}  + q_{i}^{\mu} p_{i}^{\nu}}{p_{i}.q_{i}} \quad ,
\end{equation}
where $p_{i}$ is the $i^{\rm th}$-gluon momentum, $q_{i}$ is the
corresponding reference momentum which is an arbitrary light like
4-vector and $s$ stands for spin (polarization) of gluons. We choose
$q_{1}=p_{2}$ and $q_{2}=p_{1}$ for our calculation. Finally, traces
over the Dirac matrices are carried out in $d$ dimensions.

The expressions involve thousands of three-loop scalar
integrals. However, they are expressible in terms of a much smaller
set of scalar integrals, called master integrals (MIs), by use of
integration-by-parts (IBP) \cite{Tkachov:1981wb, Chetyrkin:1981qh} and
Lorentz invariance (LI) \cite{Gehrmann:1999as} identities.  These
identities follow from the Poincare invariance of the integrands, they
result in a large linear system of equations for the integrals
relevant to given external kinematics at a fixed loop-order. The LI
identities are not linearly independent from the IBP
identities~\cite{Lee:2008tj}, their inclusion does however help to 
accelerate the solution of the system of equations. By employing
lexicographic ordering of these integrals (Laporta
algorithm,~\cite{Laporta:2001dd}), a reduction to MIs is accomplished.
Several implementations of the Laporta algorithm exist in the
literature: AIR~\cite{Anastasiou:2004vj}, FIRE~\cite{Smirnov:2008iw},
Reduze2~\cite{vonManteuffel:2012np, Studerus:2009ye} and
LiteRed~\cite{Lee:2013mka, Lee:2012cn}.  In the context of the present
calculation, we used LiteRed~\cite{Lee:2013mka, Lee:2012cn} to perform
the reductions of all the integrals to MIs.

Each three-loop Feynman integral is expressed in terms of a list of
propagators involving loop momenta that can be attributed to one of
the following three sets (auxiliary
topologies,~\cite{Gehrmann:2010ue})
\begin{align}
  \label{eq:Basis}
  {\rm A}_1 &: \{ \cD_1, \cD_2, \cD_3, \cD_{12}, \cD_{13},
              \cD_{23}, \cD_{1;1}, \cD_{1;12}, \cD_{2;1}, \cD_{2;12}, \cD_{3;1},
              \cD_{3;12} \}
              \nonumber\\
  {\rm A}_2 &: \{ \cD_1, \cD_2, \cD_3, \cD_{12}, \cD_{13}, \cD_{23},
              \cD_{13;2}, \cD_{1;12}, \cD_{2;1}, \cD_{12;2},
              \cD_{3;1}, \cD_{3;12} \}
              \nonumber\\
  {\rm A}_3 &: \{ \cD_1, \cD_2, \cD_3, \cD_{12}, \cD_{13}, \cD_{123},
              \cD_{1;1}, \cD_{1;12}, \cD_{2;1}, \cD_{2;12}, \cD_{3;1},
              \cD_{3;12} \}\, .
\end{align}
In the above sets
\begin{align*}
  \cD_{i} = k_{i}^2, \cD_{ij} = (k_i-k_j)^2, \cD_{ijl} = (k_i-k_j-k_l)^2,
\end{align*}
\vspace{-0.8cm}
\begin{align*}
  \cD_{i;j} = (k_i-p_j)^2, \cD_{i;jl} = (k_i-p_j-p_l)^2, \cD_{ij;l} = (k_i-k_j-p_l)^2 
\end{align*}
To accomplish this, we have used the package
Reduze2~\cite{vonManteuffel:2012np, Studerus:2009ye}.  In each set in
Eq.~(\ref{eq:Basis}), $\cD'$s are linearly independent and form a
complete basis in a sense that any Lorentz-invariant scalar product
involving loop momenta and external momenta can be expressed uniquely
in terms of $\cD'$s from that set.

As a result, we can express the unrenormalised form factors in terms
of 22 topologically different master integrals (MIs) which can be
broadly classified into three different types: genuine three-loop
integrals with vertex functions ($A_{t,i}$), three-loop propagator
integrals ($B_{t,i}$) and integrals which are product of one- and
two-loop integrals ($C_{t,i}$). These integrals were computed
analytically as Laurent series in $\epsilon$
in~\cite{Gehrmann:2005pd,Gehrmann:2006wg,Heinrich:2007at,Heinrich:2009be,Lee:2010cga}
and are collected in the appendix of~\cite{Gehrmann:2010ue}.
Inserting those, we obtain the final expressions for the
unrenormalised (bare) form factors that are listed in
Appendix~\ref{app:Results}.

\subsection{UV Renormalisation}
\label{ss:UV}

To obtain ultraviolet-finite expressions for the form factors, a
renormalisation of the coupling constant and of the operators is
required. The UV renormalisation of the operators
$\left[ O_{G} \right]_{B}$ and $\left[ O_{J} \right]_{B}$ involves
some non-trivial prescriptions. These are in part related to the
formalism used for the $\gamma_{5}$ matrix, section~\ref{sec:gamma5}
above.

This formalism fails to preserve the anti-commutativity of
$\gamma_{5}$ with $\gamma^{\mu}$ in $d$ dimensions. In addition, the
standard properties of the axial current and Ward identities, which
are valid in a basic regularization scheme like the one of
Pauli-Villars, are violated as well. As a consequence, one fails to
restore the correct renormalised axial current, which is defined as
\cite{Larin:1993tq, Akyeampong:1973xi}
\begin{align}
  \label{eq:J5}
  J^{\mu}_{5} \equiv \bar{\psi}\gamma^{\mu}\gamma_{5}\psi = i
  \frac{1}{3!} \varepsilon^{\mu\nu_{1}\nu_{2}\nu_{3}} \bar{\psi}
  \gamma_{\nu_{1}} \gamma_{\nu_{2}}\gamma_{\nu_{3}} \psi
\end{align}
in dimensional regularization. To rectify this, one needs to introduce
a finite renormalisation constant $Z^{s}_{5}$
\cite{Adler:1969gk,Kodaira:1979pa} in addition to the standard overall
ultraviolet renormalisation constant $Z^{s}_{\overline{MS}}$ within
the $\overline{MS}$-scheme:
\begin{align}
  \label{eq:J5Ren}
  \left[ J^{\mu}_{5} \right]_{R} = Z^{s}_{5} Z^{s}_{\overline{MS}} \left[ J^{\mu}_{5} \right]_{B}\,.
\end{align}
By evaluating the appropriate Feynman diagrams explicitly,
$Z^{s}_{\overline{MS}}$ can be computed, however the finite
renormalisation constant is not fixed through this calculation. To
determine $Z^{s}_{5}$ one has to demand the conservation of the one
loop character \cite{Adler:1969er} of the operator relation of the
axial anomaly in dimensional regularization:
\begin{align}
  \label{eq:Anomaly}
  \left[ \partial_{\mu}J^{\mu}_{5} \right]_{R} &= a_{s} \frac{n_{f}}{2} \left[ G\tilde{G} \right]_{R}
                                                 \nonumber\\
  \text{i.e.}~~~ \left[ O_{J} \right]_{R} &= a_{s} \frac{n_{f}}{2} \left[ O_{G} \right]_{R}\,.
\end{align}
The bare operator $\left[ O_{J} \right]_{B}$ is renormalised
multiplicatively exactly in the same way as the axial current
$J^{\mu}_{5}$ through
\begin{align}
  \label{eq:OJRen}
  \left[ O_{J} \right]_{R} = Z^{s}_{5} Z^{s}_{\overline{MS}} \left[ O_{J}\right]_{B}\,,
\end{align}
whereas the other one $\left[ O_{G} \right]_{B}$ mixes under the
renormalisation through
\begin{align}
  \left[ O_{G} \right]_{R} = Z_{GG} \left[ O_{G}\right]_B +
  Z_{GJ} \left[ O_{J} \right]_B
\end{align}
with the corresponding renormalisation constants $Z_{GG}$ and
$Z_{GJ}$. The above two equations can be combined to express them
through the matrix equation
\begin{align}
  \label{eq:OpMat}
  \left[ O_{i} \right]_{R} &= Z_{ij} \left[  O_{j}\right]_{B} 
\end{align}
with
\begin{align}
  \label{eq:ZMat}
  i,j &= \{G, J\}\,, 
        \nonumber\\
  O \equiv
  \begin{bmatrix}
    O_{G}\\
    O_{J}
  \end{bmatrix}
  \qquad\quad &\text{and}  \qquad\quad
                Z \equiv
                \begin{bmatrix}
                  Z_{GG} & Z_{GJ}\\
                  Z_{JG} & Z_{JJ}
                \end{bmatrix}\,.
\end{align}
In the above expressions
\begin{align}
  \label{eq:ZJGZJJ}
  Z_{JG} &= 0 \qquad \text{to all orders in perturbation theory}\,,
           \nonumber\\
  Z_{JJ} &\equiv Z^{s}_{5} Z^{s}_{\overline{MS}}\,.
\end{align}
We determine the above-mentioned renormalisation constants
$Z^{s}_{\overline{MS}},Z_{GG},Z_{GJ}$ up to
${\cal{O}}\left( a_{s}^{3} \right)$ from our calculation of the bare
on-shell pseudo-scalar form factors described in the previous
subsection. This procedure provides a completely independent approach
to their original computation, which was done in the operator product
expansion~\cite{Zoller:2013ixa}.

Our approach to compute those $Z_{ij}$ is based on the infrared
evolution equation for the form factor, and will be detailed in
Section~\ref{ss:IR} below.  Moreover, we can fix $Z^{s}_{5}$ up to
${\cal O}(a_{s}^{2})$ by demanding the operator relation of the axial
anomaly (Eq.~(\ref{eq:Anomaly})).  Using these overall operator
renormalisation constants along with strong coupling constant
renormalisation through $Z_{a_{s}}$, Eq.~(\ref{eq:Zas}), we obtain the
UV finite on-shell quark and gluon form factors.

To define the UV renormalised form factors, we introduce a quantity
${\cal{S}}^{\lambda}_{\beta}$, constructed out of bare matrix
elements, through
\begin{align}
  \label{eq:CalSG}
  {\cal{S}}^{G}_{g} &\equiv Z_{GG} \langle {\hat{\cal
                      M}}^{G,(0)}_{g}|{{\cal M}}^{G}_{g}\rangle + Z_{GJ} \langle {\hat{\cal
                      M}}^{G,(0)}_{g}|{{\cal M}}^{J}_{g}\rangle 
                      \nonumber\\
  \intertext{and}
  {\cal{S}}^{G}_{q} &\equiv Z_{GG} \langle {\hat{\cal
                      M}}^{J,(0)}_{q}|{{\cal M}}^{G}_{q}\rangle + Z_{GJ} \langle {\hat{\cal
                      M}}^{J,(0)}_{q}|{{\cal M}}^{J}_{q}\rangle \,.
\end{align}
Expanding the quantities appearing on the right hand side of the above
equation in powers of $a_{s}$ :
\begin{align}
  \label{eq:MZExpRenas}
  |{\cal M}^{\lambda}_{\beta}\rangle &= \sum_{n=0}^{\infty} {a}^{n}_{s}
                                       |{\cal M}^{\lambda,(n)}_{\beta}\rangle\,,
                                       \nonumber\\
  Z_{I} &= \sum_{n=0}^{\infty} a^{n}_{s} Z^{(n)}_{I} \qquad \text{with}
          \qquad I=GG, GJ\,\,\,,
\end{align}
we can write \\
\begin{equation}
  {\cal S}^{G}_{g} = \sum_{n=0}^{\infty} a^{n}_{s} {\cal S}^{G,(n)}_{g}\qquad
  \text{and}\qquad {\cal S}^{G}_{q} = \sum_{n=1}^{\infty} a^{n}_{s} {\cal S}^{G,(n)}_{q}\,.
  \label{eq:CalSG}
\end{equation}
\\
Then the UV renormalised form factors corresponding to $O_{G}$ are
defined as
\begin{align}
  \label{eq:RenFFG}
  \left[ {\cal F}^{G}_{g} \right]_{R} \equiv \frac{{\cal S}^{G}_{g}}{{\cal
  S}^{G,(0)}_{g}} 
  &=
    Z_{GG} {\cal
    F}^{G}_{g} + Z_{GJ} {\cal F}^{J}_{g}
    \frac{\langle {\cal
    M}^{G,(0)}_{g}|{\cal M}^{J,(1)}_{g}\rangle}{\langle {\cal
    M}^{G,(0)}_{g}|{\cal M}^{G,(0)}_{g}\rangle} 
    \nonumber\\
  &\equiv 1 + \sum^{\infty}_{n=1} a^{n}_{s} \left[ {\cal
    F}^{G,(n)}_{g} \right]_{R} \,,
    \nonumber\\ \nonumber\\
  \left[ {\cal F}^{G}_{q}
  \right]_{R}  \equiv \frac{{\cal S}^{G}_{q}}{a_{s} {\cal
  S}^{G,(1)}_{q}} 
  &= \frac{Z_{GG} {\cal F}^{G}_{q} \langle {\cal
    M}^{J,(0)}_{q}|{\cal M}^{G,(1)}_{q}\rangle + Z_{GJ} {\cal
    F}^{J}_{q} \langle {\cal
    M}^{J,(0)}_{q}|{\cal M}^{J,(0)}_{q}\rangle}{a_{s} \left[ \langle {\cal
    M}^{J,(0)}_{q}|{\cal M}^{G,(1)}_{q}\rangle + Z^{(1)}_{GJ} \langle {\cal
    M}^{J,(0)}_{q}|{\cal M}^{J,(0)}_{q}\rangle \right]} 
    \nonumber\\
  &
    \equiv 1 + \sum^{\infty}_{n=1} a^{n}_{s} 
    \left[ {\cal
    F}^{G,(n)}_{q} \right]_{R} \,
\end{align}
where
\begin{align}
  \label{eq:SGg0SGq1}
  {\cal
  S}^{G,(0)}_{g} &= \langle {\cal
                   M}^{G,(0)}_{g}|{\cal M}^{G,(0)}_{g}\rangle \,,
                   \nonumber\\
  {\cal
  S}^{G,(1)}_{q} &= 
                   \langle {\cal M}^{J,(0)}_{q}|{\cal M}^{G,(1)}_{q}\rangle 
                   + Z^{(1)}_{GJ}\langle {\cal M}^{J,(0)}_{q}|{\cal M}^{J,(0)}_{q}\rangle\,.
\end{align}
Similarly, for defining the UV finite form factors for the other
operator $O_{J}$ we introduce

\begin{align}
  \label{eq:CalSJ}
  {\cal S}^{J}_{g} &\equiv Z^{s}_{5} Z^{s}_{\overline{MS}} \langle
                     {\hat{\cal M}}^{G,(0)}_{g}| {\cal M}^{J}_{g} \rangle\,
                     \nonumber\\
  \intertext{and}
  {\cal S}^{J}_{q} &\equiv Z^{s}_{5} Z^{s}_{\overline{MS}} \langle
                     {\hat{\cal M}}^{J,(0)}_{q}| {\cal M}^{J}_{q} \rangle\,.
\end{align}
Expanding $Z^{s}_{\overline{MS}}$ and
$|{\cal{M}}^{\lambda}_{\beta}\rangle$ in powers of $a_{s}$, following
Eq.~(\ref{eq:MZExpRenas}), we get
\begin{align}
  \label{eq:CalSJExpand}  
  {\cal S}^{J}_{g} = \sum_{n=1}^{\infty} a^{n}_{s} {\cal S}^{J,(n)}_{g}
  % \intertext{and}
  \qquad\quad \text{and} \qquad\quad
  {\cal S}^{J}_{q} = \sum_{n=0}^{\infty} a^{n}_{s} {\cal S}^{J,(n)}_{q}\,.
\end{align}
\\
With these we define the UV renormalised form factors corresponding to
$O_{J}$ through
\begin{align}
  \label{eq:RenFFJ}
  \left[ {\cal F}^{J}_{g} \right]_{R} &\equiv \frac{{\cal S}^{J}_{g}}{a_{s}{\cal
                                        S}^{J,(1)}_{g}} = Z^{s}_{5}
                                        Z^{s}_{\overline{MS}} {\cal
                                        F}^{J}_{g} \equiv 1 +
                                        \sum^{\infty}_{n=1} a^{n}_{s} \left[ {\cal
                                        F}^{J,(n)}_{g} \right]_{R} \,,
                                        \nonumber\\
  \left[ {\cal F}^{J}_{q}
  \right]_{R}  &\equiv \frac{{\cal S}^{J}_{q}}{{\cal
                 S}^{J,(0)}_{q}} = Z^{s}_{5}
                 Z^{s}_{\overline{MS}} {\cal
                 F}^{J}_{q} = 1 + \sum^{\infty}_{n=1} a^{n}_{s} 
                 \left[ {\cal
                 F}^{J,(n)}_{q} \right]_{R} \,
\end{align}
where
\begin{align}
  \label{eq:SGg0SGq1}
  {\cal
  S}^{J,(1)}_{g} &= \langle {\cal
                   M}^{G,(0)}_{g}|{\cal M}^{J,(1)}_{g}\rangle \,,
                   \nonumber\\
  {\cal
  S}^{J,(0)}_{q} &= 
                   \langle {\cal M}^{J,(0)}_{q}|{\cal M}^{J,(0)}_{q}\rangle\,.
\end{align}
The finite renormalisation constant $Z^{s}_{5}$ is multiplied in
Eq.~(\ref{eq:CalSJ}) to restore the axial anomaly equation in
dimensional regularisation. We determine all required renormalisation
constants from consistency conditions on the universal structure of
the infrared poles of the renormalised form factors in the next
section, and use these constants to derive the UV-finite form factors
in Section~\ref{ss:Ren}.

\subsection{Infrared Singularities and Universal Pole Structure}
\label{ss:IR}

The renormalised form factors are ultraviolet-finite, but still
contain divergences of infrared origin.  In the calculation of
physical quantities (which fulfill certain infrared-safety
criteria~\cite{Sterman:1977wj}), these infrared singularities are
cancelled by contributions from real radiation processes that yield
the same observable final state, and by mass factorization
contributions associated with initial-state partons.  The pole
structures of these infrared divergences arising in QCD form factors
exhibit some universal behaviour. The very first successful proposal
along this direction was presented by Catani~\cite{Catani:1998bh} (see
also \cite{Sterman:2002qn}) for one and two-loop QCD amplitudes using
the universal subtraction operators. The factorization of the single
pole in quark and gluon form factors in terms of soft and collinear
anomalous dimensions was first revealed in \cite{Ravindran:2004mb} up
to two loop level whose validity at three loop was later established
in the article \cite{Moch:2005tm}. The proposal by Catani was
generalized beyond two loops by Becher and
Neubert~\cite{Becher:2009cu} and by Gardi and
Magnea~\cite{Gardi:2009qi}. Below, we outline this behaviour in the
context of pseudo-scalar form factors up to three loop level,
following closely the notation used in~\cite{Ravindran:2005vv}.

The unrenormalised form factors
${\cal F}^{\lambda}_{\beta}(\hat{a}_{s}, Q^{2}, \mu^{2}, \epsilon)$
satisfy the so-called $KG$-differential equation \cite{Sudakov:1954sw,
  Mueller:1979ih, Collins:1980ih, Sen:1981sd} which is dictated by the
factorization property, gauge and renormalisation group (RG)
invariances:
\begin{equation}
  \label{eq:KG}
  Q^2 \frac{d}{dQ^2} \ln {\cal F}^{\lambda}_{\beta} (\hat{a}_s, Q^2, \mu^2, \epsilon)
  = \frac{1}{2} \left[ K^{\lambda}_{\beta} (\hat{a}_s, \frac{\mu_R^2}{\mu^2}, \epsilon
    )  + G^{\lambda}_{\beta} (\hat{a}_s, \frac{Q^2}{\mu_R^2}, \frac{\mu_R^2}{\mu^2}, \epsilon ) \right]
\end{equation}
where all poles in the dimensional regulator $\ep$ are contained in
the $Q^{2}$ independent function $K^{\lambda}_{\beta}$ and the finite
terms in $\epsilon \rightarrow 0$ are encapsulated in
$G^{\lambda}_{\beta}$. RG invariance of the form factor implies
\begin{equation}
  \label{eq:KIA}
  \mu_R^2 \frac{d}{d\mu_R^2} K^{\lambda}_{\beta}(\hat{a}_s, \frac{\mu_R^2}{\mu^2},
  \epsilon )  = - \mu_R^2 \frac{d}{d\mu_R^2} G^{\lambda}_{\beta}(\hat{a}_s,
  \frac{Q^2}{\mu_R^2},  \frac{\mu_R^2}{\mu^2}, \epsilon ) 
  = - A^{\lambda}_{\beta} (a_s (\mu_R^2)) = - \sum_{i=1}^{\infty}  a_s^i (\mu_R^2) A^{\lambda}_{\beta,i} 
\end{equation}
where, $A^{\lambda}_{\beta,i}$ on the right hand side are the $i$-loop
cusp anomalous dimensions. It is straightforward to solve for
$K^{\lambda}_{\beta}$ in Eq.~(\ref{eq:KIA}) in powers of bare strong
coupling constant $\ashat$ by performing the following expansion
\begin{align}
  K^{\lambda}_{\beta}\l({\hat a}_{s}, \frac{\mu_{R}^{2}}{\mu^{2}}, \ep\r)  =
  \sum_{i=1}^{\infty}  {\hat a}_{s}^{i}
  \l(\frac{\mu_{R}^{2}}{\mu^{2}}\r)^{i\frac{\ep}{2}}  S_{\ep}^{i} K^{\lambda}_{\beta,i}(\ep)\, .
\end{align}
The solutions $K^{\lambda}_{\beta,i}(\ep)$ consist of simple poles in
$\ep$ with the coefficients consisting of $A_{\beta, i}^{\lambda}$ and
$\beta_{i}$. These can be found in \cite{Ravindran:2005vv,
  Ravindran:2006cg}. On the other hand, the renormalisation group
equation (RGE) of
$G^{\lambda}_{\beta,i}(\hat{a}_s, \frac{Q^2}{\mu_R^2},
\frac{\mu_R^2}{\mu^2}, \epsilon )$
can be solved. The solution contains two parts, one is dependent on
$\mu_{R}^{2}$ whereas the other part depends only the boundary point
$\mu^{2}_{R}=Q^{2}$. The $\mu_{R}^{2}$ dependent part can eventually
be expressed in terms of $A^{\lambda}_{\beta}$:
\begin{align}
  \label{eq:GSoln}
  G^{\lambda}_{\beta}(\hat{a}_s, \frac{Q^2}{\mu_R^2},
  \frac{\mu_R^2}{\mu^2}, \epsilon ) = G^{\lambda}_{\beta}({a}_s(Q^{2}),
  1, \epsilon ) + \int_{\frac{Q^{2}}{\mu_{R}^{2}}}^{1} \frac{dx}{x}  
  A^{\lambda}_{\beta}(a_{s}\l(x\mu_{R}^{2})\r)\,. 
\end{align}
The boundary term can be expanded in powers of $a_{s}$ as
\begin{align}
  G^{\lambda}_{\beta}(a_s(Q^2), 1, \epsilon) = \sum_{i=1}^{\infty} a_s^i(Q^2) G^{\lambda}_{\beta,i}(\epsilon)\, .
\end{align}
The solutions of $K^{\lambda}_{\beta}$ and $G^{\lambda}_{\beta}$
enable us to solve the $KG$ equation (Eq.~(\ref{eq:KG})) and thereby
facilitate to obtain the
$\ln {\cal F}^{\lambda}_{\beta}(\hat{a}_s, Q^2, \mu^2, \ep)$ in terms
of $A^{\lambda}_{\beta, i}, G^{\lambda}_{\beta, i}$ and $\beta_{i}$
which is given by~\cite{Ravindran:2005vv}
\begin{align}
  \label{eq:lnFSoln}
  \ln {\cal F}^{\lambda}_{\beta}(\hat{a}_s, Q^2, \mu^2, \ep) =
  \sum_{i=1}^{\infty} {\hat a}_{s}^{i} \l(\frac{Q^{2}}{\mu^{2}}\r)^{i
  \frac{\ep}{2}} S_{\ep}^{i} \hat {\cal L}_{\beta,i}^{\lambda}(\ep)
\end{align}
with
\begin{align}
  \label{eq:lnFitoCalLF}
  \hat {\cal L}_{\beta,1}^{\lambda}(\ep) =& { \frac{1}{\ep^2} } \Bigg\{-2 A^{\lambda}_{\beta,1}\Bigg\}
                                            + { \frac{1}{\ep}
                                            }
                                            \Bigg\{G^{\lambda}_{\beta,1}
                                            (\ep)\Bigg\}\, ,
                                            \nonumber\\
  \hat {\cal L}_{\beta,2}^{\lambda}(\ep) =& { \frac{1}{\ep^3} } \Bigg\{\beta_0 A^{\lambda}_{\beta,1}\Bigg\}
                                            + {
                                            \frac{1}{\ep^2} }
                                            \Bigg\{-  {
                                            \frac{1}{2} }  A^{\lambda}_{\beta,2}
                                            - \beta_0   G^{\lambda}_{\beta,1}(\ep)\Bigg\}
                                            + { \frac{1}{\ep}
                                            } \Bigg\{ {
                                            \frac{1}{2} }  G^{\lambda}_{\beta,2}(\ep)\Bigg\}\, ,
                                            \nonumber\\
  \hat {\cal L}_{\beta,3}^{\lambda}(\ep) =& { \frac{1}{\ep^4}
                                            } \Bigg\{- {
                                            \frac{8}{9} }  \beta_0^2 A^{\lambda}_{\beta,1}\Bigg\}
                                            + {
                                            \frac{1}{\ep^3} }
                                            \Bigg\{ { \frac{2}{9} } \beta_1 A^{\lambda}_{\beta,1}
                                            + { \frac{8}{9} }
                                            \beta_0 A^{\lambda}_{\beta,2}  + { \frac{4}{3} }
                                            \beta_0^2 G^{\lambda}_{\beta,1}(\ep)\Bigg\}
                                            \nonumber\\
                                          &
                                            + { \frac{1}{\ep^2} } \Bigg\{- { \frac{2}{9} } A^{\lambda}_{\beta,3}
                                            - { \frac{1}{3} } \beta_1 G^{\lambda}_{\beta,1}(\ep)
                                            - { \frac{4}{3} } \beta_0 G^{\lambda}_{\beta,2}(\ep)\Bigg\}
                                            + { \frac{1}{\ep}
                                            } \Bigg\{  { \frac{1}{3} } G^{\lambda}_{\beta,3}(\ep)\Bigg\}\, .
\end{align}
All these form factors are observed to satisfy \cite{Ravindran:2004mb,
  Moch:2005tm} the following decomposition in terms of collinear
($B^{\lambda}_{\beta}$), soft ($f^{\lambda}_{\beta}$) and UV
($\gamma^{\lambda}_{\beta}$) anomalous dimensions:
\begin{align}
  \label{eq:GIi}
  G^{\lambda}_{\beta,i} (\ep) = 2 \left(B^{\lambda}_{\beta,i} -
  \gamma^{\lambda}_{\beta,i}\right)  + f^{\lambda}_{\beta,i} +
  C^{\lambda}_{\beta,i}  + \sum_{k=1}^{\infty} \epsilon^k g^{\lambda,k}_{\beta,i} \, ,
\end{align}
where the constants $C^{\lambda}_{\beta,i}$ are given by
\cite{Ravindran:2006cg}
\begin{align}
  \label{eq:Cg}
  C^{\lambda}_{\beta,1} &= 0\, ,
                          \nonumber\\
  C^{\lambda}_{\beta,2} &= - 2 \beta_{0} g^{\lambda,1}_{\beta,1}\, ,
                          \nonumber\\
  C^{\lambda}_{\beta,3} &= - 2 \beta_{1} g^{\lambda,1}_{\beta,1} - 2
                          \beta_{0} \left(g^{\lambda,1}_{\beta,2}  + 2 \beta_{0} g^{\lambda,2}_{\beta,1}\right)\, .
\end{align}
In the above expressions, $X^{\lambda}_{\beta,i}$ with $X=A,B,f$ and
$\gamma^{\lambda}_{\beta, i}$ are defined through
\begin{align}
  \label{eq:ABfgmExp}
  X^{\lambda}_{\beta} &\equiv \sum_{i=1}^{\infty} a_{s}^{i}
                        X^{\lambda}_{\beta,i}\,,
                        \qquad \text{and} \qquad
                        \gamma^{\lambda}_{\beta} \equiv \sum_{i=1}^{\infty} a_{s}^{i} \gamma^{\lambda}_{\beta,i}\,\,.
\end{align}
Within this framework, we will now determine this universal structure
of IR singularities of the pseudo-scalar form factors. This
prescription will be used subsequently to determine the overall
operator renormalisation constants.

We begin with the discussion of form factors corresponding to
$O_{J}$. The results of the form factors ${\cal F}^{J}_{\beta}$ for
$\beta=q,g$, which have been computed up to three loop level in this
article are being used to extract the unknown factors,
$\gamma^{J}_{\beta,i}$ and $g^{J,k}_{\beta,i}$, by employing the $KG$
equation. Since the ${\cal F}^{J}_{\beta}$ satisfy $KG$ equation, we
can obtain the solutions Eq.~(\ref{eq:lnFSoln}) along with
Eq.~(\ref{eq:lnFitoCalLF}) and Eq.~(\ref{eq:GIi}) to examine our
results against the well known decomposition of the form factors in
terms of the quantities $X^{J}_{\beta}$. These are universal, and
appear also in the vector and scalar quark and gluon form
factors~\cite{Moch:2005tm, Ravindran:2004mb}. They are
known~\cite{Vogt:2004mw, Catani:1990rp, Vogt:2000ci, Ravindran:2004mb,
  Ahmed:2014cha} up to three loop level in the literature.  Using
these in the above decomposition, we obtain
$\gamma^{J}_{\beta,i}$. The other process dependent constants, namely,
$g^{J,k}_{\beta,i}$ can be obtained by comparing the coefficients of
$\epsilon^{k}$ in Eq.~(\ref{eq:lnFitoCalLF}) at every order in
${\hat a}_{s}$. We can get the quantities $\gamma^{J}_{g,i}$ and
$g^{J,k}_{g,i}$ up to two loop level, since this process starts at one
loop. From gluon form factors we get
\begin{align}
  \label{eq:gmJqQ}
  \gamma^{J}_{g,1} &= 0\,,
                     \nonumber\\
  \gamma^{J}_{g,2} &= {\dis{C_{A} C_{F}}} \Bigg\{- \frac{44}{3} \Bigg\}+
                     {\dis{C_{F} n_{f}}} \Bigg\{- \frac{10}{3}
                     \Bigg\}\,. \hspace{6cm}
\end{align}
Similarly, from the quark form factors we obtain
\begin{align}
  \label{eq:gmJqQ}
  \gamma^{J}_{q,1} &= 0\,,
                     \nonumber\\
  \gamma^{J}_{q,2} &= {\dis{C_{A} C_{F}}} \Bigg\{- \frac{44}{3} \Bigg\}+
                     {\dis{C_{F} n_{f}}} \Bigg\{- \frac{10}{3} \Bigg\}\,, 
                     \nonumber\\
  \gamma^{J}_{q,3} &= {\dis{C^{2}_{A} C_{F}}} \Bigg\{ -
                     \frac{3578}{27}\Bigg\} +
                     {\dis{C^{2}_{F}n_{f}}} \Bigg\{\frac{22}{3}\Bigg\}
                     - {\dis{C_{F} n^{2}_{f}}}
                     \Bigg\{\frac{26}{27}\Bigg\} + {\dis{C_{A}
                     C^{2}_{F}}} \Bigg\{\frac{308}{3}\Bigg\} 
                     \nonumber\\
                   &+
                     {\dis{C_{A} C_{F} n_{f}}} \Bigg\{-\frac{149}{27}\Bigg\}\,.
\end{align}
Note that $\gamma^{J}_{q,i} = \gamma^{J}_{g,i}$ which is expected
since these are the UV anomalous dimensions associated with the same
operator $[O_{J}]_{B}$. The $\gamma^{J}_{\beta,i}$ are further used to
obtain the overall operator renormalisation constant
$Z^{s}_{\overline{MS}}$ through the RGE:
\begin{align}
  \label{eq:RGEZMS}
  \mu_{R}^{2}\frac{d}{d\mu_{R}^{2}}
  \ln Z^{\lambda}(a_{s},\mu_{R}^{2},\epsilon) = \sum_{i=1}^{\infty}
  a_{s}^{i} \gamma^{\lambda}_{i}.
\end{align}
\\
The general solution of the RGE is obtained as
\begin{align}
  \label{eq:GenSolZI}
  Z^{\lambda} &= 1 + a_s \Bigg[ \frac{1}{\epsilon} 2  {\gamma^{\lambda}_{1}} \Bigg]  +
                a_s^2 \Bigg[  \frac{1}{\epsilon^2} \Bigg\{ 2 \beta_0
                {\gamma^{\lambda}_{1}}  + 2 ({\gamma^{\lambda}_{1}})^2 \Bigg\} + 
                \frac{1}{\epsilon} {\gamma^{\lambda}_{2}}  \Bigg] 
                + a_s^3 \Bigg[
                \frac{1}{\epsilon^{3}} \Bigg\{ 8 \beta_0^2 {\gamma^{\lambda}_{1}}  +
                4 \beta_0 ({\gamma^{\lambda}_{1}})^2 
                \nonumber\\
              &+
                \frac{4 ({\gamma^{\lambda}_{1}})^3}{3} \Bigg\} +  \frac{1}{\epsilon^{2}} \Bigg\{ \frac{4
                \beta_1 {\gamma^{\lambda}_{1}}}{3} + \frac{4 \beta_0
                {\gamma^{\lambda}_{2}}}{3}  + 2 {\gamma^{\lambda}_{1}}
                {\gamma^{\lambda}_{2}} \Bigg\} + 
                \frac{1}{\epsilon} \Bigg\{ \frac{2 {\gamma^{\lambda}_{3}}}{3} \Bigg\}
                \Bigg]\,.
\end{align}
By substituting the results of $\gamma^{J}_{\beta,i}$ in the above
solution we get $Z^{s}_{\overline{MS}}$ up to ${\cal O}(a_{s}^{3})$:

\begin{align}
  \label{eq:ZMS}
  Z^{s}_{\overline{MS}} &= 1 + a^{2}_{s} \Bigg[C_{A} C_{F} \Bigg\{-
                          \frac{44}{3 \epsilon} \Bigg\} + C_{F} n_{f} \Bigg\{ -
                          \frac{10}{3 \epsilon} \Bigg\}
                          \Bigg]
                          + a^{3}_{s} \Bigg[ C_{A}^2 C_{F} \Bigg\{ -
                          \frac{1936}{27 \epsilon^2} - \frac{7156}{81 \epsilon}
                          \Bigg\}
                          \nonumber\\
                        &+  C_{F}^2 n_{f} \Bigg\{ \frac{44}{9 \epsilon} \Bigg\} + 
                          C_{F} n_{f}^2 \Bigg\{ \frac{80}{27 \epsilon^2} - \frac{52}{81 \epsilon} \Bigg\}  + 
                          C_{A}  C_{F}^2 \Bigg\{ \frac{616}{9 \epsilon}\Bigg\} +
                          C_{A} C_{F} n_{f} \Bigg\{ - \frac{88}{27 \epsilon^2} -
                          \frac{298}{81 \epsilon} \Bigg\}
                          \Bigg],
\end{align}
which agrees completely with the known result in \cite{Larin:1993tq}.
In order to restore the axial anomaly equation in dimensional
regularization (see Section~\ref{ss:UV} above), we must multiply the
$Z^{s}_{\overline{MS}} \left[ O_{J} \right]_{B}$ by a finite
renormalisation constant $Z^{s}_{5}$, which reads \cite{Larin:1993tq}
\begin{align}
  \label{eq:Z5s}
  Z^{s}_{5} = 1 + a_{s} \{-4 C_{F}\} + a^{2}_{s} \left\{ 22 C^{2}_{F} -
  \frac{107}{9} C_{A} C_{F} + \frac{31}{18} C_{F} n_{f} \right\}\,.
\end{align}
Following the computation of the operator mixing constants below, we
will be able to verify explicitly that this expression yields the
correct expression for the axial anomaly.

Now, we move towards the discussion of $O_{G}$ form factors. Similar
to previous case, we consider the form factors
$Z_{GG}^{-1} [ {\cal F}^{G}_{\beta} ]_{R}$, defined through
Eq.~(\ref{eq:RenFFG}), to extract the unknown constants,
$\gamma^{G}_{\beta,i}$ and $g^{G,k}_{\beta,i}$, by utilizing the $KG$
differential equation. Since, $[{\cal F}^G_{\beta}]_{R}$ is UV finite,
the product of $Z_{GG}^{-1}$ with $[{\cal F}^G_{\beta}]_{R}$ can
effectively be treated as unrenormalised form factor and hence we can
demand that $Z_{GG}^{-1} [{\cal F}^{G}_{\beta}]_{R}$ satisfy $KG$
equation.  Further we make use of the solutions Eq.~(\ref{eq:lnFSoln})
in conjunction with Eq.~(\ref{eq:lnFitoCalLF}) and Eq.~(\ref{eq:GIi})
to compare our results against the universal decomposition of the form
factors in terms of the constants $X^{G}_{\beta}$. Upon substituting
the existing results of the quantities
$A^{G}_{\beta,i}, B^{G}_{\beta,i}$ and $f^{G}_{\beta,i}$ up to three
loops, which are obtained in case of quark and gluon form factors, we
determine the anomalous dimensions $\gamma^{G}_{\beta,i}$ and the
constants $g^{G,k}_{\beta,i}$. However, it is only possible to get the
factors $\gamma^{G}_{q,i}$ and $g^{G,k}_{q,i}$ up to two loops because
of the absence of a tree level amplitude in the quark initiated
process for the operator $O_{G}$. Since $[{\cal F}^G_{\beta}]_{R}$ are
UV finite, the anomalous dimensions $\gamma^{G}_{\beta,i}$ must be
equal to the anomalous dimension corresponding to the renormalisation
constant $Z_{GG}$. This fact is being used to determine the overall
renormalisation constants $Z_{GG}$ and $Z_{GJ}$ up to three loop level
where these quantities are parameterized in terms of the newly
introduced anomalous dimensions $\gamma_{ij}$ through the matrix
equation
\begin{align}
  \label{eq:ZijDefn}
  \mu_{R}^{2}\frac{d}{d\mu_{R}^{2}}Z_{ij} \equiv \gamma_{ik} Z_{kj}\,
  \qquad \text{with} \qquad i,j,k={G,J}
\end{align}
The general solution of the RGE up to $a_{s}^{3}$ is obtained as
\begin{align}
  \label{eq:ZCoupSoln}
  Z_{ij} &= \delta_{ij} 
           + {a}_{s} \Bigg[ \frac{2}{\epsilon}
           \gamma_{ij,1} \Bigg] 
           + {a}_{s}^{2} \Bigg[
           \frac{1}{\epsilon^{2}} \Bigg\{  2
           \beta_{0} \gamma_{ij,1} + 2 \gamma_{ik,1} \gamma_{kj,1}  \Bigg\} + \frac{1}{\epsilon} \Bigg\{ \gamma_{ij,2}\Bigg\}
           \Bigg] 
           + {a}_{s}^{3} \Bigg[ \frac{1}{\epsilon^{3}} \Bigg\{
           \frac{8}{3} \beta_{0}^{2} \gamma_{ij,1} 
           \nonumber\\
         &+ 4 \beta_{0} \gamma_{ik,1}
           \gamma_{kj,1} + \frac{4}{3} \gamma_{ik,1} \gamma_{kl,1}
           \gamma_{lj,1} \Bigg\} + \frac{1}{\epsilon^{2}} \Bigg\{ \frac{4}{3} \beta_{1} \gamma_{ij,1} +
           \frac{4}{3} \beta_{0} \gamma_{ij,2} 
           + \frac{2}{3}
           \gamma_{ik,1} \gamma_{kj,2} 
           \nonumber\\
         &+ \frac{4}{3}
           \gamma_{ik,2} \gamma_{kj,1} \Bigg\} + \frac{1}{\epsilon}
           \Bigg\{ \frac{2}{3} \gamma_{ij,3} \Bigg\} \Bigg]
\end{align}
where, $\gamma_{ij}$ is expanded in powers of $a_{s}$ as
\begin{align}
  \label{eq:gammaijExp}
  \gamma_{ij} = \sum_{n=1}^{\infty} a_{s}^{n} \gamma_{ij,n}\,.
\end{align}
Demanding the vanishing of $\gamma^{G}_{\beta,i}$, we get
\begin{align}
  \label{eq:gammaGG}
  \gamma_{GG} &= a_{s} \Bigg[ \frac{11}{3} C_{A} - \frac{2}{3} n_{f}\Bigg] +
                a_{s}^{2} \Bigg[ \frac{34}{3} C_{A}^{2} - \frac{10}{3} C_{A} n_{f} -
                2 C_{F} n_{f} \Bigg]
                + a_{s}^{3} \Bigg[ \frac{2857}{54} C_{A}^3 - \frac{1415}{54} C_{A}^2
                n_{f}  
                \nonumber\\
              &- \frac{205}{18} C_{A} C_{F} n_{f} + C_{F}^2 n_{f} +
                \frac{79}{54} C_{A} n_{f}^2 + \frac{11}{9} C_{F} n_{f}^2\Bigg]\,,
                \nonumber\\
  \gamma_{GJ} &= a_{s} \Bigg[ - 12 C_{F} \Bigg] + a_{s}^{2} \Bigg[
                - \frac{284}{3} C_{A} C_{F}  + 36 C_{F}^2 +
                \frac{8}{3} C_{F} n_{f} \Bigg]
                + a_{s}^{3} \Bigg[ - \frac{1607}{3} C_{A}^2 C_{F} 
                \nonumber\\
              &+ 461 C_{A}  C_{F}^2 - 126 C_{F}^3  - \frac{164}{3} C_{A} C_{F} n_{f} + 
                214 C_{F}^2 n_{f} + \frac{52}{3} C_{F} n_{f}^2
                + 288 C_{A} C_{F}
                n_{f} \zeta_3  
                \nonumber\\
              &- 288 C_{F}^2 n_{f} \zeta_3
                \Bigg]\,.
\end{align}
In addition to the demand of vanishing $\gamma^{G}_{\beta,i}$, it is
required to use the results of $\gamma_{JJ}$ and $\gamma_{JG}$, which
are implied by the definition, Eq.~(\ref{eq:ZijDefn}), up to
${\cal O}(a_{s}^{2})$ to determine the above-mentioned $\gamma_{GG}$
and $\gamma_{GJ}$ up to the given order. This is a consequence of the
fact that the operators mix under UV renormalisation. Following
Eq.~(\ref{eq:ZijDefn}) along with Eq.~(\ref{eq:ZJGZJJ}),
Eq.~(\ref{eq:ZMS}) and Eq.~(\ref{eq:Z5s}), we obtain
\begin{align}
  \label{eq:gammaJJJG}
  \gamma_{JJ} &= a_{s} \Bigg[ - \epsilon 2 C_{F}  \Bigg] + a_{s}^{2}
                \Bigg[ \epsilon \Bigg\{ - \frac{107}{9} C_A C_F + 14
                C_F^2 + \frac{31}{18} C_F n_f
                \Bigg\} - 6
                C_F n_f  \Bigg]
                \intertext{and}
                \gamma_{JG} &=0\,.
\end{align}
As it happens, we note that $\gamma_{JJ}$'s are $\epsilon$-dependent
and in fact, this plays a crucial role in determining the other
quantities. Our results are in accordance with the existing ones,
$\gamma_{GG}$ and $\gamma_{GJ}$, which are available up to
${\cal O}(a_{s}^{2})$ \cite{Larin:1993tq} and ${\cal O}(a_{s}^{3})$
\cite{Zoller:2013ixa}, respectively. In addition to the existing ones,
here we compute the new result of $\gamma_{GG}$ at
${\cal O}(a_{s}^{3})$. It was observed through explicit computation in
the article \cite{Larin:1993tq} that
\begin{align}
  \label{eq:gammaGGbt}
  \gamma_{GG} = - \frac{\beta}{a_{s}} 
\end{align} 
holds true up to two loop level but there was no statement on the
validity of this relation beyond that order. In \cite{Zoller:2013ixa},
it was demonstrated in the operator product expansion that the
relation holds even at three loop. Here, through explicit calculation,
we arrive at the same conclusion that the relation is still valid at
three loop level which can be seen if we look at the $\gamma_{GG, 3}$
in Eq.~(\ref{eq:gammaGG}) which is equal to the $\beta_{2}$.

Before ending the discussion of $\gamma_{ij}$, we examine our results
against the axial anomaly relation. The renormalisation group
invariance of the anomaly equation (Eq.~(\ref{eq:Anomaly})), see
\cite{Larin:1993tq}, gives
\begin{align}
  \label{eq:AnomalyAlt}
  \gamma_{JJ} = \frac{\beta}{a_{s}} + \gamma_{G{G}} + a_{s}
  \frac{n_{f}}{2} \gamma_{GJ}\,.
\end{align}
Through our calculation up to three loop level we find that our
results are in complete agreement with the above anomaly equation
through
\begin{align}
  \label{eq:AnomalyHold}
  \gamma_{GG} &= - \frac{\beta}{a_{s}}
                \qquad \text{and} \qquad
                \gamma_{GJ} = \left( a_{s} \frac{n_{f}}{2} \right)^{-1} \gamma_{JJ}
\end{align}
in the limit of $\epsilon \rightarrow 0$. This serves as one of the
most crucial checks on our computation.

Additionally, if we conjecture the above relations to hold beyond
three loops (which could be doubted in light of recent
findings~\cite{Almelid:2015jia}), then we can even predict the
$\epsilon$-independent part of the $\gamma_{JJ}$ at
${\cal O}(a_{s}^{3})$:
\begin{align}
  \label{eq:gammaJJ4}
  \gamma_{JJ}|_{\epsilon \rightarrow 0} &=  a_{s}^{2} \Bigg[ - 6 C_{F} n_{f} \Bigg] + a_{s}^{3} \Bigg[
                                          - \frac{142}{3} C_{A} C_{F} n_{f}  + 18 C_{F}^2 n_{f} +
                                          \frac{4}{3} C_{F} n_{f}^{2} \Bigg]\,.
\end{align}

The results of $\gamma_{ij}$ uniquely specify $Z_{ij}$, through
Eq.~(\ref{eq:ZCoupSoln}).  We summarize the resulting expressions of
$Z_{ij}$ below:
\begin{align}
  \label{eq:ZGGtZGJ}
  Z_{GG} &= 1 +  a_s \Bigg[ \frac{22}{3\epsilon}
           C_{A}  -
           \frac{4}{3\epsilon} n_{f} \Bigg] 
           + 
           a_s^2 \Bigg[ \frac{1}{\epsilon^2}
           \Bigg\{ \frac{484}{9} C_{A}^2 - \frac{176}{9} C_{A}
           n_{f} + \frac{16}{9} n_{f}^2 \Bigg\}
           + \frac{1}{\epsilon} \Bigg\{ \frac{34}{3} C_{A}^2  
           \nonumber\\
         &-
           \frac{10}{3} C_{A} n_{f}  - 2 C_{F} n_{f} \Bigg\} \Bigg] 
           + 
           a_s^3 \Bigg[   \frac{1}{\epsilon^3} 
           \Bigg\{ \frac{10648}{27} C_{A}^3 - \frac{1936}{9}
           C_{A}^2 n_{f}  + \frac{352}{9} C_{A} n_{f}^2  -
           \frac{64}{27} n_{f}^3 \Bigg\}  
           \nonumber\\
         &+   \frac{1}{\epsilon^2}
           \Bigg\{ \frac{5236}{27} C_{A}^3 - \frac{2492}{27}
           C_{A}^2 n_{f}  - \frac{308}{9} C_{A} C_{F} n_{f}  + 
           \frac{280}{27} C_{A} n_{f}^2  + \frac{56}{9} C_{F}
           n_{f}^2 \Bigg\}
           \nonumber\\
         &  
           +  \frac{1}{\epsilon} \Bigg\{ \frac{2857}{81} C_{A}^3  -
           \frac{1415}{81} C_{A}^2 n_{f}  - \frac{205}{27} C_{A} C_{F} n_{f} + 
           \frac{2}{3} C_{F}^2 n_{f} + \frac{79}{81} C_{A}
           n_{f}^2  + \frac{22}{27} C_{F} n_{f}^2 \Bigg\}
           \Bigg]
           \nonumber \intertext{and} 
           Z_{GJ} &=  a_s \Bigg[ - \frac{24}{\epsilon} C_{F} \Bigg]
                    + 
                    a_s^2 \Bigg[ \frac{1}{\epsilon^2}
                    \Bigg\{ - 176 C_{A} C_{F} + 32 C_{F} n_{f} \Bigg\}
                    + \frac{1}{\epsilon} \Bigg\{ - \frac{284}{3} C_{A} C_{F} +
                    84 C_{F}^2 
                    \nonumber\\
         &+ \frac{8}{3} C_{F} n_{f} \Bigg\}  \Bigg]
           + a_{s}^{3} \Bigg[
           \frac{1}{\epsilon^3} \Bigg\{ - \frac{3872}{3} C_{A}^2 C_{F}  +
           \frac{1408}{3} C_{A} C_{F} n_{f}  - \frac{128}{3} C_{F}
           n_{f}^2 \Bigg\}  
           \nonumber\\
         &+ 
           \frac{1}{\epsilon^2} \Bigg\{ - \frac{9512}{9} C_{A}^2 C_{F}  +
           \frac{2200}{3} C_{A} C_{F}^2  + \frac{2272}{9} C_{A} C_{F}
           n_{f}  - 
           \frac{64}{3} C_{F}^2 n_{f} - \frac{32}{9} C_{F} n_{f}^2 \Bigg\}
           \nonumber\\
         &+ 
           \frac{1}{\epsilon} \Bigg\{ - \frac{3214}{9} C_{A}^2 C_{F}  +
           \frac{5894}{9} C_{A} C_{F}^2  - 356 C_{F}^3 - \frac{328}{9}
           C_{A} C_{F} n_{f}  + 
           \frac{1096}{9} C_{F}^2 n_{f} + \frac{104}{9} C_{F} n_{f}^2 
           \nonumber\\
         &+ 192
           C_{A} C_{F} n_{f} \zeta_3  - 
           192 C_{F}^2 n_{f} \zeta_3 \Bigg\} 
           \Bigg]\,.
\end{align}
$Z_{GG}$ and $Z_{GJ}$ are in agreement with the results already
available in the literature up to ${\cal{O}}(a^{2}_{s})$
\cite{Larin:1993tq} and ${\cal{O}}(a_{s}^{3})$ \cite{Zoller:2013ixa},
where a completely different approach and methodology was used.

\subsection{Results of UV Renormalised Form Factors}
\label{ss:Ren}

Using the renormalisation constants obtained in the previous section,
we get all the UV renormalised form factors
$\left[ {\cal F}^{\lambda}_{\beta} \right]_{R}$, defined in
Eq.~(\ref{eq:RenFFG}) and Eq.~(\ref{eq:RenFFJ}), up to three loops. In
this section we present the results for the choice of the scales
$\mu_R^2=\mu_F^2=q^2$.
\begin{align}
  \label{eq:FFRen1Ggg}
  \left[ {\cal F}^{G,(1)}_{g} \right]_{R} &=  {\dis{2 n_{f} T_{F}}} \Bigg\{ - \frac{4}{3
                                            \epsilon} \Bigg\}  + {\dis{C_{A}}} \Bigg\{ -
                                            \frac{8}{\epsilon^2}  +
                                            \frac{22}{3 \epsilon} + 4 + \zeta_2  + \epsilon \Bigg( - 6 -
                                            \frac{7}{3} \zeta_3 \Bigg) + 
                                            \epsilon^2 \Bigg( 7 -
                                            \frac{\zeta_2}{2}  
                                            \nonumber\\
                                          &+ \frac{47}{80} \zeta_2^2
                                            \Bigg) + 
                                            \epsilon^3 \Bigg( - \frac{15}{2}
                                            + \frac{3}{4} \zeta_2   +
                                            \frac{7}{6} \zeta_3  +
                                            \frac{7}{24} \zeta_2 \zeta_3  - 
                                            \frac{31}{20} \zeta_5 \Bigg) \Bigg\}\,,
                                            % \end{align}
                                            % %
                                            % \begin{align}
                                            %   \label{eq:FFRen2Ggg}
  \\  \left[ {\cal F}^{G,(2)}_{g} \right]_{R} &=  {\dis{4 n_{f}^2 T_{F}^{2}}} \Bigg\{ \frac{16}{9
                                                \epsilon^2} \Bigg\}
                                                + 
                                                {\dis{C_{A}^2}} \Bigg\{
                                                \frac{32}{\epsilon^4}  -
                                                \frac{308}{3 \epsilon^3}  + 
                                                \Bigg( \frac{62}{9} - 4 \zeta_2
                                                \Bigg) \frac{1}{\epsilon^2}  + 
                                                \Bigg( \frac{2780}{27} +
                                                \frac{11}{3} \zeta_2  +
                                                \frac{50}{3} \zeta_3 \Bigg)
                                                \frac{1}{\epsilon} 
                                                \nonumber\\
                                          &-
                                            \frac{3293}{81} +
                                            \frac{115}{6} \zeta_2  -
                                            \frac{21}{5} \zeta_2^2  - 33 \zeta_3   + 
                                            \epsilon \Bigg( -
                                            \frac{114025}{972} -
                                            \frac{235}{18} \zeta_2  +
                                            \frac{1111}{120} \zeta_2^2  +
                                            \frac{1103}{54} \zeta_3  
                                            \nonumber\\
                                          &- 
                                            \frac{23}{6} \zeta_2 \zeta_3 -
                                            \frac{71}{10} \zeta_5 \Bigg) + 
                                            \epsilon^2 \Bigg(
                                            \frac{4819705}{11664} -
                                            \frac{694}{27} \zeta_2  -
                                            \frac{2183}{240} \zeta_2^2  + 
                                            \frac{2313}{280} \zeta_2^3 -
                                            \frac{7450}{81} \zeta_3  
                                            \nonumber\\
                                          &-
                                            \frac{11}{36} \zeta_2 \zeta_3  + 
                                            \frac{901}{36} \zeta_3^2 -
                                            \frac{341}{20} \zeta_5 \Bigg)  
                                            \Bigg\}  
                                            + 
                                            {\dis{2 C_{A} n_{f} T_{F}}} \Bigg\{ \frac{56}{3
                                            \epsilon^3}  - \frac{52}{3
                                            \epsilon^2}  + \Bigg( -
                                            \frac{272}{27} -  \frac{2}{3}
                                            \zeta_2 \Bigg) \frac{1}{\epsilon}
                                            \nonumber\\
                                          & - \frac{295}{81} 
                                           - 
                                            \frac{5}{3} \zeta_2 - 2 \zeta_3 +
                                            \epsilon \Bigg( \frac{15035}{486}
                                            + \frac{\zeta_2}{18} +
                                            \frac{59}{60} \zeta_2^2  + 
                                            \frac{383}{27} \zeta_3 \Bigg) +
                                            \epsilon^2 \Bigg( -
                                            \frac{116987}{1458}  +
                                            \frac{583}{108} \zeta_2  
                                            \nonumber\\
                                          &- 
                                            \frac{329}{72} \zeta_2^2 -
                                            \frac{1688}{81} \zeta_3  +
                                            \frac{61}{18} \zeta_2 \zeta_3  - 
                                            \frac{49}{10} \zeta_5 \Bigg)
                                            \Bigg\} 
                                            + 
                                            {\dis{2 C_{F} n_{f} T_{F}}} \Bigg\{  - \frac{2}{\epsilon} -
                                            \frac{71}{3} + 8
                                            \zeta_3 + 
                                            \epsilon \Bigg( \frac{2665}{36} 
                                            \nonumber\\
                                          &-
                                            \frac{19}{6} \zeta_2  -
                                            \frac{8}{3} \zeta_2^2  -
                                            \frac{64}{3} \zeta_3 \Bigg) + 
                                            \epsilon^2 \Bigg( -
                                            \frac{68309}{432} +
                                            \frac{505}{36} \zeta_2  +
                                            \frac{64}{9} \zeta_2^2 +
                                            \frac{455}{9} \zeta_3 -  
                                            \frac{10}{3} \zeta_2 \zeta_3  
                                            \nonumber\\
                                          &+  8 \zeta_5 \Bigg) \Bigg\}\,,
                                          % \end{align}
                                          % %
                                          % \begin{align}
                                          %   \label{eq:FFRen3Ggg}
  \\  \left[ {\cal F}^{G,(3)}_{g} \right]_{R} &= {\dis{8 n_{f}^3 T_{F}^{3}}} \Bigg\{ -  
                                                \frac{64}{27 \epsilon^3} \Bigg\}  
                                                +
                                                {\dis{4 C_{F} n_{f}^2 T_{F}^{2}}} \Bigg\{ \frac{56}{9
                                                \epsilon^2} + \Bigg(
                                                \frac{874}{27}  - \frac{32}{3}
                                                \zeta_3 \Bigg) \frac{1}{\epsilon}
                                                -
                                                \frac{418}{27} + 2 \zeta_2  + 
                                                \frac{16}{5} \zeta_2^2 
                                                \nonumber\\
                                          &- \frac{80}{9} \zeta_3 \Bigg\} 
                                            + 
                                            {\dis{2 C_{F}^2 n_{f} T_{F}}} \Bigg\{
                                            \frac{2}{3
                                            \epsilon} + \frac{457}{6} + 104 \zeta_3  - 160
                                            \zeta_5 \Bigg\} 
                                            + 
                                            {\dis{2 C_{A}^2 n_{f} T_{F}}} \Bigg\{ -
                                            \frac{320}{3 \epsilon^5}  
\nonumber\\
&+
                                            \frac{28480}{81 \epsilon^4} 
                                           +
                                            \Bigg( - \frac{608}{243}  +
                                            \frac{56}{27} \zeta_2 \Bigg)
                                            \frac{1}{\epsilon^3}  + 
                                            \Bigg( - \frac{54088}{243} +
                                            \frac{676}{81} \zeta_2  +
                                            \frac{272}{27}  \zeta_3 \Bigg)
                                            \frac{1}{\epsilon^2} 
\nonumber\\
&+ 
                                            \Bigg( - \frac{623293}{2187} 
                                           -
                                            \frac{7072}{243} \zeta_2 -
                                            \frac{941}{90} \zeta_2^2  -
                                            \frac{7948}{81}  \zeta_3 \Bigg)
                                            \frac{1}{\epsilon} +
                                            \frac{6345979}{13122} - 
                                            \frac{42971}{729} \zeta_2 +
                                            \frac{687}{20} \zeta_2^2 
\nonumber\\
&+
                                            \frac{652}{3} \zeta_3  
                                           -
                                            \frac{301}{9} \zeta_2 \zeta_3   +
                                            \frac{4516}{45} \zeta_5 \Bigg\}  
                                            + 
                                            {\dis{4 C_{A} n_{f}^2 T_{F}^{2}}} \Bigg\{ -
                                            \frac{2720}{81 \epsilon^4}  +
                                            \frac{7984}{243 \epsilon^3}  + 
                                            \Bigg( \frac{560}{27} 
\nonumber\\
&+
                                            \frac{8}{27} \zeta_2 \Bigg)
                                            \frac{1}{\epsilon^2} 
                                           + \Bigg(
                                            \frac{10889}{2187}  +
                                            \frac{140}{81} \zeta_2  +
                                            \frac{328}{81} \zeta_3 \Bigg)
                                            \frac{1}{\epsilon} + 
                                            \frac{9515}{6561} +
                                            \frac{10}{27} \zeta_2  -
                                            \frac{157}{135} \zeta_2^2  - 
                                            \frac{20}{243} \zeta_3  \Bigg\}  
                                            \nonumber\\
                                          &+ 
                                            {\dis{2 C_{A} C_{F} n_{f} T_{F}}} \Bigg\{
                                            \frac{272}{9 \epsilon^3}   + 
                                            \Bigg( \frac{4408}{27} -
                                            \frac{640}{9} \zeta_3 \Bigg)
                                            \frac{1}{\epsilon^2}  + 
                                            \Bigg( - \frac{65110}{81} +
                                            \frac{74}{3} \zeta_2  +
                                            \frac{352}{15} \zeta_2^2 
                                            \nonumber\\
                                          &+
                                            \frac{6496}{27} \zeta_3 \Bigg)
                                            \frac{1}{\epsilon} + 
                                            \frac{1053625}{972} -
                                            \frac{311}{2} \zeta_2  -
                                            \frac{1168}{15} \zeta_2^2 -
                                            \frac{24874}{81} \zeta_3 +  48 \zeta_2 \zeta_3 + 
                                            \frac{32}{9} \zeta_5 \Bigg\}
                                            \nonumber\\
                                          &+ 
                                            {\dis{C_{A}^3}}  \Bigg\{  -
                                            \frac{256}{3 \epsilon^6}  + 
                                            \frac{1760}{3 \epsilon^5} -
                                            \frac{62264}{81 \epsilon^4}  + 
                                            \Bigg( - \frac{176036}{243} -
                                            \frac{308}{27} \zeta_2  -
                                            \frac{176}{3} \zeta_3 \Bigg)
                                            \frac{1}{\epsilon^3} + 
                                            \Bigg( \frac{207316}{243} 
                                            \nonumber\\
                                          &-
                                            \frac{8164}{81} \zeta_2  +
                                            \frac{494}{45} \zeta_2^2  +
                                            \frac{9064}{27} \zeta_3  \Bigg)
                                            \frac{1}{\epsilon^2} +
                                            \Bigg( \frac{2763800}{2187}  +
                                            \frac{36535}{243} \zeta_2  - 
                                            \frac{12881}{180} \zeta_2^2 -
                                            \frac{3988}{9} \zeta_3  
                                            \nonumber\\
                                          &+
                                            \frac{170}{9} \zeta_2 \zeta_3  + 
                                            \frac{1756}{15} \zeta_5 \Bigg)
                                            \frac{1}{\epsilon} -
                                            \frac{84406405}{26244} +
                                            \frac{617773}{1458} \zeta_2  + 
                                            \frac{144863}{1080} \zeta_2^2 -
                                            \frac{22523}{270} \zeta_2^3   
                                            \nonumber\\
                                          &+
                                            \frac{44765}{243} \zeta_3 -  
                                            \frac{1441}{18} \zeta_2 \zeta_3 -
                                            \frac{1766}{9} \zeta_3^2 +
                                            \frac{13882}{45} \zeta_5 \Bigg\}\,,
                                            % \end{align}
                                            % %
                                            % \begin{align}
                                            %   \label{eq:FFRen1GqQ}
  \\  \left[ {\cal F}^{G,(1)}_{q} \right]_{R} &= {\dis{C_{F}}} \Bigg\{ -
                                                \frac{8}{\epsilon^2}  +
                                                \frac{6}{\epsilon}  -
                                                \frac{33}{4} + \zeta_2  + 
                                                \epsilon \Bigg( \frac{29}{16}
                                                + \frac{25}{48}
                                                \zeta_2  - \frac{7}{3}
                                                \zeta_3 \Bigg)  + 
                                                \epsilon^2 \Bigg( \frac{299}{192}
                                                - \frac{1327}{576} \zeta_2 
                                                \nonumber\\
                                          &+
                                            \frac{1387}{2880} \zeta_2^2  + 
                                            \frac{143}{48} \zeta_3 \Bigg) + \epsilon^3 \Bigg( -
                                            \frac{13763}{2304}  +
                                            \frac{32095}{6912} \zeta_2  - 
                                            \frac{1559}{3456} \zeta_2^2 + \frac{61}{6912} \zeta_2^3 -
                                            \frac{1625}{576} \zeta_3  
                                            \nonumber\\
                                          &+ 
                                            \frac{377}{864} \zeta_2 \zeta_3 - \frac{31}{20} \zeta_5 \Bigg)
                                            \Bigg\}  
                                            + 
                                            {2 \dis{n_{f} T_{F}}} \Bigg\{ -
                                            \frac{445}{162}  + \epsilon \Bigg( \frac{8231}{1944}  -
                                            \frac{239}{1944} \zeta_2  -
                                            \frac{2}{3} \zeta_3 \Bigg) 
                                            \nonumber\\
                                          &+ 
                                            \epsilon^2 \Bigg( -
                                            \frac{50533}{7776}  + \frac{1835}{7776} \zeta_2
                                            + \frac{22903}{116640} \zeta_2^2 + 
                                            \frac{9125}{5832} \zeta_3  +
                                            \frac{1}{18} \zeta_2 \zeta_3 \Bigg)  + 
                                            \epsilon^3 \Bigg(
                                            \frac{2754151}{279936}  
                                            \nonumber\\
                                          &- \frac{35083}{93312}
                                            \zeta_2  - \frac{316343}{699840} \zeta_2^2 - 
                                            \frac{22903}{1399680} \zeta_2^3
                                            - \frac{61121}{23328}  \zeta_3  + 
                                            \frac{2053}{34992} \zeta_2
                                            \zeta_3  - \frac{1}{216} \zeta_2^2
                                            \zeta_3  
                                            \nonumber\\
                                          &- \frac{7}{54} \zeta_3^2 - 
                                            \frac{7}{6} \zeta_5 \Bigg) \Bigg\}
                                            + 
                                            {\dis{C_{A}}} \Bigg\{ \frac{7115}{324}
                                            - \frac{2}{3} \zeta_2 - 2 \zeta_3 + 
                                            \epsilon \Bigg( -
                                            \frac{114241}{3888}  + \frac{7321}{3888} \zeta_2
                                            + \frac{53}{90} \zeta_2^2   
                                            \nonumber\\
                                          &+ \frac{13}{3} \zeta_3 + 
                                            \frac{1}{6} \zeta_2 \zeta_3 \Bigg)
                                            + \epsilon^2 \Bigg(
                                            \frac{692435}{15552}  -
                                            \frac{55117}{15552} \zeta_2  - 
                                            \frac{326369}{233280} \zeta_2^2
                                            - \frac{53}{1080} \zeta_2^3  -
                                            \frac{90235}{11664} \zeta_3  
                                            \nonumber\\
                                          &- 
                                            \frac{41}{108} \zeta_2 \zeta_3 -
                                            \frac{1}{72} \zeta_2^2 \zeta_3  -
                                            \frac{7}{18} \zeta_3^2  - 5 \zeta_5
                                            \Bigg)  + 
                                            \epsilon^3 \Bigg( - \frac{37171073}{559872} +
                                            \frac{1013165}{186624} \zeta_2  
                                            \nonumber\\
                                          &+ 
                                            \frac{3399073}{1399680} \zeta_2^2
                                            +  \frac{34037663}{19595520}
                                            \zeta_2^3  + 
                                            \frac{53}{12960} \zeta_2^4 +
                                            \frac{585439}{46656} \zeta_3   -
                                            \frac{56159}{69984} \zeta_2 \zeta_3
                                            + 
                                            \frac{3223}{12960} \zeta_2^2
                                            \zeta_3 
                                            \nonumber\\
                                          &+ \frac{1}{864} \zeta_2^3 \zeta_3 + \frac{8}{9}
                                            \zeta_3^2  + 
                                            \frac{7}{108} \zeta_2 \zeta_3^2 +
                                            8 \zeta_5   + \frac{5}{12} \zeta_2 \zeta_5 \Bigg) \Bigg\}\,,
                                            % \end{align}
                                            % %
                                            % \begin{align}
                                            %   \label{eq:FFRen2GqQ}
  \\  \left[ {\cal F}^{G,(2)}_{q} \right]_{R} &= {\dis{4 n_{f}^2 T_{F}^{2}}}
                                                \Bigg\{
                                                \frac{9505}{1458}  +
                                                \epsilon \Bigg(  -
                                                \frac{146177}{5832}  +
                                                \frac{12419}{17496}
                                                \zeta_2 + \frac{38}{9}
                                                \zeta_3 \Bigg) \Bigg\}
                                                + 
                                                {\dis{2 C_{F} n_{f} T_{F}}} \Bigg\{
                                                \frac{8}{\epsilon^3}  +
                                                \frac{1636}{81 \epsilon^2}  
                                                \nonumber\\
                                          &+
                                            \Bigg( - \frac{12821}{243}  -
                                            \frac{247}{243} \zeta_2   +
                                            \frac{16}{3} \zeta_3 \Bigg)
                                            \frac{1}{\epsilon} + 
                                            \frac{20765}{324} + \frac{35}{486} \zeta_2 +
                                            \frac{85}{2916} \zeta_2^2  +
                                            \frac{6265}{729} \zeta_3  - 
                                            \frac{4}{9} \zeta_2 \zeta_3  
                                            \nonumber\\
                                          &+ 
                                            \epsilon \Bigg(-
                                            \frac{1457425}{34992}  -
                                            \frac{11146}{729} \zeta_2 -
                                            \frac{232457}{174960}  \zeta_2^2 - 
                                            \frac{85}{34992} \zeta_2^3
                                            + \frac{9907}{1458} \zeta_3 -
                                            \frac{7723}{4374} \zeta_2 \zeta_3
                                            \nonumber\\
                                          &+ 
                                            \frac{1}{27} \zeta_2^2
                                            \zeta_3  +
                                            \frac{28}{27} \zeta_3^2   - \frac{20}{9} \zeta_5
                                            \Bigg) \Bigg\} 
                                            + 
                                            {\dis{C_{A}^2}} \Bigg\{
                                            \frac{2796445}{5832}  -
                                            \frac{587}{18} \zeta_2  + 
                                            \frac{53}{30} \zeta_2^2 
                                            -
                                            \frac{185}{2} \zeta_3  -
                                            \frac{10}{3} \zeta_2 \zeta_3  
                                            \nonumber\\
                                          &+
                                            20 \zeta_5   + 
                                            \epsilon \Bigg( -
                                            \frac{34321157}{23328}  +
                                            \frac{10420379}{69984} \zeta_2  +
                                            \frac{589}{20} \zeta_2^2  + 
                                            \frac{7921}{2520} \zeta_2^3 
                                            +
                                            \frac{8411}{24} \zeta_3  -
                                            \frac{329}{72} \zeta_2 \zeta_3  
                                            \nonumber\\
                                          &+ 
                                            \frac{5}{18} \zeta_2^2 \zeta_3 +
                                            13 \zeta_3^2  - \frac{757}{18} \zeta_5 - 
                                            \frac{5}{3} \zeta_2 \zeta_5
                                            \Bigg)  \Bigg\}
                                            + 
                                            {\dis{2 C_{A} n_{f} T_{F}}} \Bigg\{ -
                                            \frac{178361}{1458}  + \frac{44}{9} \zeta_2 -
                                            \frac{76}{45} \zeta_2^2  
\nonumber\\
&-
                                            \frac{44}{9} \zeta_3  
                                           + 
                                            \epsilon \Bigg(
                                            \frac{2357551}{5832}  -
                                            \frac{478171}{17496} \zeta_2  -
                                            \frac{137}{135} \zeta_2^2  + 
                                            \frac{19}{135} \zeta_2^3 -
                                            \frac{1621}{27} \zeta_3  -
                                            \frac{40}{27} \zeta_2
  \zeta_3  
\nonumber\\
&+ 
                                            \frac{22}{3} \zeta_5 \Bigg)  
                                            \Bigg\} 
                                           + 
                                            {\dis{C_{A} C_{F}}} \Bigg\{  -
                                            \frac{44}{\epsilon^3}  
                                            + 
                                            \Bigg( - \frac{13654}{81} +
                                            \frac{28}{3} \zeta_2  + 16
                                            \zeta_3 \Bigg)
                                            \frac{1}{\epsilon^2} + 
                                            \Bigg( \frac{186925}{486} 
\nonumber\\
&-
                                            \frac{3919}{486} \zeta_2  -
                                            \frac{212}{45} \zeta_2^2  
                                           -
                                            \frac{218}{3} \zeta_3  
                                            - 
                                            \frac{4}{3} \zeta_2 \zeta_3
                                            \Bigg) \frac{1}{\epsilon} -
                                            \frac{61613}{81}  +
                                            \frac{59399}{972} \zeta_2  + 
                                            \frac{749513}{29160}
  \zeta_2^2 
\nonumber\\
&+
                                            \frac{53}{135} \zeta_2^3
                                            + \frac{213517}{1458} \zeta_3 + 
                                            \frac{91}{27} \zeta_2 \zeta_3 
                                           +
                                            \frac{1}{9} \zeta_2^2
                                            \zeta_3  + \frac{28}{9} \zeta_3^2  +
                                            \epsilon \Bigg(
                                            \frac{35327209}{34992}  -
                                            \frac{2158003}{23328}
                                            \zeta_2 
\nonumber\\
&-
                                            \frac{3532645}{69984}
                                            \zeta_2^2  - 
                                            \frac{11307767}{2449440}
                                            \zeta_2^3  -
                                            \frac{53}{1620} \zeta_2^4
                                           - \frac{1030169}{2916} \zeta_3 + 
                                            \frac{191915}{8748}
                                            \zeta_2 \zeta_3  -
                                            \frac{817}{405} \zeta_2^2
                                            \zeta_3  
\nonumber\\
&- \frac{1}{108} \zeta_2^3
                                            \zeta_3  - 
                                            \frac{121}{9} \zeta_3^2 -
                                            \frac{14}{27}  \zeta_2
                                            \zeta_3^2  - \frac{43}{6} \zeta_5
                                            \Bigg)  \Bigg\} 
                                           + 
                                            {\dis{C_{F}^2}} \Bigg\{
                                            \frac{32}{\epsilon^4}  -
                                            \frac{48}{\epsilon^3}   + 
                                            \Bigg( 84 - 8 \zeta_2 \Bigg)
                                            \frac{1}{\epsilon^2}  
                                            \nonumber\\
&+ \Bigg(
                                            - \frac{125}{2}  - \frac{61}{6}
                                            \zeta_2  + \frac{128}{3} \zeta_3
                                            \Bigg) \frac{1}{\epsilon} +
                                            \frac{6881}{216} +
                                            \frac{193}{12} \zeta_2  
                                           -
                                            \frac{281}{24} \zeta_2^2  - 
                                            \frac{1037}{18} \zeta_3  
\nonumber\\
&+ 
                                            \epsilon \Bigg(
                                            \frac{166499}{2592} -
                                            \frac{3761}{648} \zeta_2
                                            + \frac{3451}{480}
                                            \zeta_2^2 - \frac{31}{288}
                                            \zeta_2^3 + 
                                            \frac{10607}{108} \zeta_3
                                            - \frac{1081}{108}
                                            \zeta_2 \zeta_3  
                                           + \frac{328}{45} \zeta_5
                                            \Bigg)  
                                            \Bigg\}\,,
                                          % \end{align}
                                          % %
                                          % \begin{align}
                                          %   \label{eq:FFRen1Jgg}
  \\  \left[ {\cal F}^{J,(1)}_{g} \right]_{R} &= {\dis{2 n_{f} T_{F}}}  \Bigg\{ - \frac{4}{3
                                                \epsilon} \Bigg\}
                                                + 
                                                {\dis{C_{A}}} \Bigg\{ -
                                                \frac{8}{\epsilon^2}  +
                                                \frac{22}{3 \epsilon} + 4 + \zeta_2  + 
                                                \epsilon \Bigg( - \frac{15}{2} +
                                                \zeta_2  - \frac{16}{3} \zeta_3
                                                \Bigg) 
\nonumber\\
&+ 
                                                \epsilon^2 \Bigg( \frac{287}{24}
                                           - 2 \zeta_2  + \frac{127}{80}
                                            \zeta_2^2 \Bigg)    + 
                                            \epsilon^3 \Bigg( -
                                            \frac{5239}{288}  +
                                            \frac{151}{48} \zeta_2  +
                                            \frac{19}{120} \zeta_2^2  +
                                            \frac{\zeta_3}{12}  + 
                                            \frac{7}{6} \zeta_2
  \zeta_3  
\nonumber\\
&-
                                            \frac{91}{20} \zeta_5 \Bigg)
                                            \Bigg\}  
                                           + 
                                            {\dis{C_{F}}} \Bigg\{  
                                            \epsilon \Bigg( - \frac{21}{2} +
                                            6 \zeta_3 \Bigg)  + \epsilon^2 \Bigg(
                                            \frac{155}{8} - \frac{5}{2}
                                            \zeta_2  - \frac{9}{5} \zeta_2^2
                                            - \frac{9}{2} \zeta_3
  \Bigg) 
\nonumber\\
&+ \epsilon^3
                                            \Bigg( - \frac{1025}{32}  +
                                            \frac{83}{16} \zeta_2  
                                           + 
                                            \frac{27}{20} \zeta_2^2 +
                                            \frac{20}{3} \zeta_3  -
                                            \frac{3}{4} \zeta_2 \zeta_3  + \frac{21}{2} \zeta_5 \Bigg) \Bigg\}\,,
                                            % \end{align}
                                            % %
                                            % \begin{align}
                                            %   \label{eq:FFRen2Jgg}
  \\  \left[ {\cal F}^{J,(2)}_{g} \right]_{R} &= {\dis{4 n_{f}^2 T_{F}^{2}}} \Bigg\{ \frac{16}{9 \epsilon^2} \Bigg\}
                                                + 
                                                {\dis{C_{A} C_{F}}} \Bigg\{\Bigg( 84 - 48
                                                \zeta_3 \Bigg) \frac{1}{\epsilon}
                                                - 232 +
                                                20 \zeta_2  + \frac{72}{5}
                                                \zeta_2^2 + 80 \zeta_3 
\nonumber\\
&+ 
                                                \epsilon \Bigg( \frac{17545}{108}
                                           - 58 \zeta_2 - 24 \zeta_2^2  -
                                            \frac{38}{3} \zeta_3  + 10
                                            \zeta_2 \zeta_3  - 
                                            14 \zeta_5 \Bigg) + 
                                            \epsilon^2 \Bigg(
                                            \frac{402635}{1296} -
                                            \frac{233}{36} \zeta_2  
\nonumber\\
&+
                                            \frac{72}{5} \zeta_2^2  +
                                            \frac{17}{70} \zeta_2^3  
                                           + 
                                            \frac{535}{12} \zeta_3 - 2
                                            \zeta_2 \zeta_3 - 34 \zeta_3^2  -
                                            \frac{1355}{6} \zeta_5 \Bigg) \Bigg\}
                                            + {\dis{2 C_{A} n_{f} T_{F}}}
                                            \Bigg\{
                                            \frac{56}{3 \epsilon^3}  -
                                            \frac{52}{3 \epsilon^2}  
\nonumber\\
&+ 
                                            \Bigg( - \frac{272}{27} -
                                            \frac{2}{3} \zeta_2 \Bigg)
                                            \frac{1}{\epsilon}   
                                           - \frac{133}{81} - 3 \zeta_2 +
                                            2 \zeta_3  + 
                                            \epsilon \Bigg( \frac{7153}{243}
                                            - \frac{7}{18} \zeta_2 -
                                            \frac{13}{60} \zeta_2^2  +
                                            \frac{599}{27} \zeta_3
  \Bigg)  
\nonumber\\
&+ 
                                            \epsilon^2 \Bigg( -
                                            \frac{135239}{1458} 
                                           +
                                            \frac{1139}{108} \zeta_2  -
                                            \frac{167}{24} \zeta_2^2  - 
                                            \frac{3146}{81} \zeta_3 +
                                            \frac{73}{18} \zeta_2 \zeta_3  -
                                            \frac{137}{30}  \zeta_5 \Bigg)
                                            \Bigg\} 
                                            \nonumber\\
&+ 
                                            {\dis{2 C_{F} n_{f} T_{F}}} \Bigg\{
                                            - \frac{2}{\epsilon}  -
                                            \frac{29}{3} 
                                           +  \epsilon
                                            \Bigg( \frac{14989}{216} -
                                            \frac{25}{6} \zeta_2 -
                                            \frac{4}{15} \zeta_2^2  - 
                                            32 \zeta_3 \Bigg) + \epsilon^2
                                            \Bigg( -
  \frac{606661}{2592}  
\nonumber\\
&+
                                            \frac{2233}{72} \zeta_2  +
                                            \frac{158}{15} \zeta_2^2  
                                           + 
                                            \frac{1409}{18} \zeta_3 - 2
                                            \zeta_2 \zeta_3  + \frac{82}{3}
                                            \zeta_5 \Bigg) \Bigg\}  
                                            + 
                                            {\dis{C_{A}^2}} \Bigg\{
                                            + \frac{32}{\epsilon^4}  -
                                            \frac{308}{3 \epsilon^3}  
\nonumber\\
&+
                                            \Bigg( \frac{62}{9}  - 4 \zeta_2
                                            \Bigg) \frac{1}{\epsilon^2} + 
                                            \Bigg( \frac{3104}{27} 
                                           -
                                            \frac{13}{3} \zeta_2  +
                                            \frac{122}{3} \zeta_3  \Bigg)
                                            \frac{1}{\epsilon}  - \frac{7397}{81} + 
                                            \frac{77}{2} \zeta_2 -
                                            \frac{61}{5} \zeta_2^2 - 55
                                            \zeta_3   
\nonumber\\
&+ 
                                            \epsilon \Bigg( -
                                            \frac{32269}{972} -
                                            \frac{997}{36} \zeta_2  
                                           +
                                            \frac{1049}{120} \zeta_2^2  -
                                            \frac{2393}{108} \zeta_3  - 
                                            \frac{53}{6} \zeta_2 \zeta_3 +
                                            \frac{369}{10} \zeta_5
  \Bigg)  
\nonumber\\
&+ 
                                            \epsilon^2 \Bigg(
                                            \frac{4569955}{11664} -
                                            \frac{15323}{432} \zeta_2  +
                                            \frac{2129}{180} \zeta_2^2  
                                           - 
                                            \frac{7591}{840} \zeta_2^3 -
                                            \frac{4099}{1296} \zeta_3  -
                                            \frac{605}{36} \zeta_2 \zeta_3  + 
                                            \frac{775}{36} \zeta_3^2 
\nonumber\\
&+
                                            \frac{2011}{30} \zeta_5 \Bigg)
                                            \Bigg\}  
                                            + 
                                            {\dis{C_{F}^2}} \Bigg\{ \epsilon \Bigg(
                                            \frac{763}{12} +  17 \zeta_3 
                                           - 60
                                            \zeta_5 \Bigg)  + 
                                            \epsilon^2 \Bigg( -
                                            \frac{18857}{144} + \frac{31}{3}
                                            \zeta_2 -  \frac{76}{15}
                                            \zeta_2^2 
\nonumber\\
&+ \frac{120}{7}
                                            \zeta_2^3  - 
                                            145 \zeta_3 + 4 \zeta_2 \zeta_3 +
                                            30 \zeta_3^2 
                                           + \frac{470}{3} \zeta_5 \Bigg) \Bigg\}\,,
                                          % \end{align}
                                          % %
                                          % \begin{align}
                                          %   \label{eq:FFRen1JqQ}
  \\  \left[ {\cal F}^{J,(1)}_{q} \right]_{R} &= {\dis{C_{F}}} \Bigg\{ -
                                                \frac{8}{\epsilon^2}  +
                                                \frac{6}{\epsilon}   - 6 + \zeta_2 +
                                                \epsilon \Bigg( - 1  -
                                                \frac{3}{4} \zeta_2   \frac{7}{3}
                                                \zeta_3 \Bigg)  + 
                                                \epsilon^2 \Bigg( \frac{5}{2} +
                                                \frac{\zeta_2}{4}  +
                                                \frac{47}{80} \zeta_2^2  +
                                                \frac{7}{4} \zeta_3 \Bigg)  
                                                \nonumber\\
                                          &+ 
                                            \epsilon^3 \Bigg( - \frac{13}{4}
                                            + \frac{\zeta_2}{8}  -
                                            \frac{141}{320} \zeta_2^2  -
                                            \frac{7}{12} \zeta_3  + 
                                            \frac{7}{24} \zeta_2 \zeta_3  -
                                            \frac{31}{20} \zeta_5  \Bigg) \Bigg\}\,,
                                            % \end{align}
                                            % %
                                            % \begin{align}
                                            %   \label{eq:FFRen2JqQ}
  \\  \left[ {\cal F}^{J,(2)}_{q} \right]_{R} &= {\dis{2 C_{F} n_{f} T_{F}}} \Bigg\{ 
                                                \frac{8}{\epsilon^3}  -
                                                \frac{16}{9 \epsilon^2}  + \Bigg(
                                                - \frac{65}{27}  - 2 \zeta_2
                                                \Bigg) \frac{1}{\epsilon}  -
                                                \frac{3115}{324} + 
                                                \frac{23}{9} \zeta_2 +
                                                \frac{2}{9} \zeta_3  + \epsilon
                                                \Bigg( \frac{129577}{3888}  
                                                \nonumber\\
                                          &-
                                            \frac{731}{108} \zeta_2  - 
                                            \frac{\zeta_2^2}{10} +
                                            \frac{119}{27} \zeta_3 \Bigg)  + 
                                            \epsilon^2 \Bigg( -
                                            \frac{3054337}{46656} +
                                            \frac{20951}{1296} \zeta_2  -
                                            \frac{145}{144} \zeta_2^2  - 
                                            \frac{2303}{324} \zeta_3 
                                            \nonumber\\
                                          &-
                                            \frac{10}{9} \zeta_2 \zeta_3  -
                                            \frac{59}{30} \zeta_5 \Bigg)
                                            \Bigg\}  
                                            + 
                                            {\dis{C_{F}^2}} \Bigg\{
                                            \frac{32}{\epsilon^4}  -
                                            \frac{48}{\epsilon^3} + \Bigg( 66
                                            - 8 \zeta_2 \Bigg)
                                            \frac{1}{\epsilon^2}   +
                                            \Bigg( - \frac{53}{2}  +
                                            \frac{128}{3} \zeta_3  \Bigg)
                                            \frac{1}{\epsilon} 
                                            \nonumber\\
                                          &- \frac{121}{8} +
                                            \frac{\zeta_2}{2}  - 
                                            13 \zeta_2^2 - 58 \zeta_3 +
                                            \epsilon \Bigg( \frac{3403}{32}
                                            + \frac{27}{8} \zeta_2  +
                                            \frac{171}{10} \zeta_2^2  + 
                                            \frac{559}{6} \zeta_3 -
                                            \frac{56}{3} \zeta_2 \zeta_3  
                                            \nonumber\\
                                          &+
                                            \frac{92}{5} \zeta_5 \Bigg) + 
                                            \epsilon^2 \Bigg( -
                                            \frac{21537}{128} -
                                            \frac{825}{32} \zeta_2  -
                                            \frac{457}{16} \zeta_2^2  +
                                            \frac{223}{20} \zeta_2^3  - 
                                            \frac{4205}{24} \zeta_3 +
                                            \frac{27}{2} \zeta_2
                                            \zeta_3  
\nonumber\\
&+
                                            \frac{652}{9} \zeta_3^2  
                                           - 
                                            \frac{231}{10} \zeta_5 \Bigg) 
                                            \Bigg\}  
                                            + 
                                            {\dis{C_{A} C_{F}}} \Bigg\{ -
                                            \frac{44}{\epsilon^3} + 
                                            \Bigg( \frac{64}{9} + 4 \zeta_2
                                            \Bigg) \frac{1}{\epsilon^2}  +
                                            \Bigg( \frac{961}{54}  + 11
                                            \zeta_2 
\nonumber\\
&- 26 \zeta_3 \Bigg)
                                            \frac{1}{\epsilon}  
                                           -
                                            \frac{30493}{648} -
                                            \frac{193}{18} \zeta_2  +
                                            \frac{44}{5} \zeta_2^2  + 
                                            \frac{313}{9} \zeta_3 + \epsilon
                                            \Bigg( - \frac{79403}{7776}  +
                                            \frac{133}{216} \zeta_2  -
                                            \frac{229}{20} \zeta_2^2  
\nonumber\\
&- 
                                            \frac{4165}{54} \zeta_3 
                                           +
                                            \frac{89}{6} \zeta_2 \zeta_3  -
                                            \frac{51}{2} \zeta_5 \Bigg)  + 
                                            \epsilon^2 \Bigg(
                                            \frac{9732323}{93312} +
                                            \frac{41363}{2592} \zeta_2  +
                                            \frac{33151}{1440} \zeta_2^2  - 
                                            \frac{809}{280} \zeta_2^3 
\nonumber\\
&+
                                            \frac{89929}{648} \zeta_3  
                                            -
                                            \frac{80}{9} \zeta_2 \zeta_3  - 
                                            \frac{569}{12} \zeta_3^2 +
                                            \frac{2809}{60}  \zeta_5 \Bigg) \Bigg\}\,,
                                            % \end{align}
                                            % %
                                            % \begin{align}
                                            %   \label{eq:FFRen3JqQ}
  \\  \left[ {\cal F}^{J,(3)}_{q} \right]_{R} &= Z^{s,(3)}_{5} 
                                                + {\dis{4 C_{F}
                                                n_{f}^2 T_{F}^{2}}} \Bigg\{ -
                                                \frac{704}{81 \epsilon^4}  +
                                                \frac{64}{243 \epsilon^3}  +
                                                \Bigg( \frac{184}{81}  +
                                                \frac{16}{9} \zeta_2 \Bigg)
                                                \frac{1}{\epsilon^2}    + 
                                                \Bigg( - \frac{4834}{2187} +
                                                \frac{40}{27} \zeta_2  
                                                \nonumber\\
                                          &+
                                            \frac{16}{81} \zeta_3 \Bigg)
                                            \frac{1}{\epsilon}  + 
                                            \frac{538231}{13122} - 
                                            \frac{680}{81} \zeta_2 -
                                            \frac{188}{135} \zeta_2^2-
                                            \frac{416}{243} \zeta_3 \Bigg\}  
                                            + 
                                            {\dis{C_{F}^3}} \Bigg\{ -
                                            \frac{256}{3 \epsilon^6}  +
                                            \frac{192}{\epsilon^5}    
                                            \nonumber\\
                                          &+
                                            \Bigg( - 336  + 32 \zeta_2 \Bigg)
                                            \frac{1}{\epsilon^4}  + 
                                            \Bigg( 280 + 24 \zeta_2 -
                                            \frac{800}{3}  \zeta_3 \Bigg)
                                            \frac{1}{\epsilon^3} + \Bigg(
                                            - 58 - 66 \zeta_2  +
                                            \frac{426}{5} \zeta_2^2  
                                            \nonumber\\
                                          &+ 552
                                            \zeta_3 \Bigg)
                                            \frac{1}{\epsilon^2}  + 
                                            \Bigg( - \frac{4193}{6} + 83
                                            \zeta_2 - \frac{1461}{10}
                                            \zeta_2^2  - \frac{3142}{3} \zeta_3 + 
                                            \frac{428}{3} \zeta_2 \zeta_3 -
                                            \frac{1288}{5}  \zeta_5 \Bigg)
                                            \frac{1}{\epsilon} 
                                            \nonumber\\
                                          &+ \frac{41395}{24} +
                                            \frac{1933}{12} \zeta_2  + 
                                            \frac{10739}{40} \zeta_2^2 -
                                            \frac{9095}{252} \zeta_2^3 + 1385
                                            \zeta_3 - 35 \zeta_2 \zeta_3  - 
                                            \frac{1826}{3} \zeta_3^2  -
                                            \frac{562}{5} \zeta_5 \Bigg\}  
                                            \nonumber\\
                                          &+ 
                                            {\dis{2 C_{F}^2 n_{f} T_{F}}} \Bigg\{ -
                                            \frac{64}{\epsilon^5}  +
                                            \frac{560}{9 \epsilon^4}  +
                                            \Bigg( - \frac{680}{27}  + 24
                                            \zeta_2 \Bigg)
                                            \frac{1}{\epsilon^3} + 
                                            \Bigg( \frac{5180}{81} -
                                            \frac{266}{9} \zeta_2  -
                                            \frac{440}{9} \zeta_3  \Bigg)
                                            \frac{1}{\epsilon^2}  
                                            \nonumber\\
                                          &+ 
                                            \Bigg( - \frac{78863}{243} +
                                            \frac{2381}{27} \zeta_2  +
                                            \frac{287}{18} \zeta_2^2  -
                                            \frac{938}{27} \zeta_3 \Bigg)
                                            \frac{1}{\epsilon}   + 
                                            \frac{1369027}{1458} -
                                            \frac{16610}{81} \zeta_2  - 
                                            \frac{8503}{1080} \zeta_2^2 
                                            \nonumber\\
                                          &+ 
                                            \frac{22601}{81} \zeta_3 +
                                            \frac{35}{3} \zeta_2 \zeta_3  -
                                            \frac{386}{9} \zeta_5 \Bigg\}  
                                            + 
                                            {\dis{C_{A}^2 C_{F}}} \Bigg\{ -
                                            \frac{21296}{81 \epsilon^4} +
                                            \Bigg( - \frac{22928}{243}  +
                                            \frac{880}{27} \zeta_2  \Bigg)
                                            \frac{1}{\epsilon^3}  
                                            \nonumber\\
                                          &+ \Bigg(
                                            \frac{23338}{243}  +
                                            \frac{6500}{81} \zeta_2  -
                                            \frac{352}{45} \zeta_2^2  - 
                                            \frac{3608}{27} \zeta_3 \Bigg)
                                            \frac{1}{\epsilon^2}    + 
                                            \Bigg( \frac{139345}{4374} +
                                            \frac{14326}{243} \zeta_2  +
                                            \frac{332}{15} \zeta_2^2 
                                            \nonumber\\
                                          & -
                                            \frac{7052}{27} \zeta_3  + 
                                            \frac{176}{9} \zeta_2 \zeta_3  +
                                            \frac{272}{3} \zeta_5  \Bigg)
                                            \frac{1}{\epsilon}  -
                                            \frac{10659797}{52488} -
                                            \frac{207547}{729} \zeta_2  + 
                                            \frac{19349}{270} \zeta_2^2 -
                                            \frac{6152}{189} \zeta_2^3 
                                            \nonumber\\
                                          &+
                                            \frac{361879}{486} \zeta_3  +
                                            \frac{344}{3} \zeta_2 \zeta_3  - 
                                            \frac{1136}{9} \zeta_3^2 -
                                            \frac{2594}{9} \zeta_5  \Bigg\}  
                                            + 
                                            {\dis{2 C_{A} C_{F} n_{f} T_{F}}} \Bigg\{ +
                                            \frac{7744}{81 \epsilon^4}  +
                                            \Bigg( \frac{6016}{243}  
                                            \nonumber\\
                                          &-
                                            \frac{160}{27} \zeta_2  \Bigg)
                                            \frac{1}{\epsilon^3} + \Bigg( -
                                            \frac{8272}{243}  -
                                            \frac{1904}{81} \zeta_2  +
                                            \frac{848}{27} \zeta_3 \Bigg)
                                            \frac{1}{\epsilon^2}   +
                                            \Bigg( \frac{17318}{2187}  -
                                            \frac{5188}{243} \zeta_2  -
                                            \frac{88}{15} \zeta_2^2  
                                            \nonumber\\
                                          &+ 
                                            \frac{1928}{81} \zeta_3 \Bigg)
                                            \frac{1}{\epsilon}    -
                                            \frac{4158659}{13122} +
                                            \frac{81778}{729} \zeta_2  -
                                            \frac{17}{135} \zeta_2^2  -
                                            \frac{5881}{27} \zeta_3  + 
                                            \frac{22}{3} \zeta_2 \zeta_3 +
                                            \frac{176}{3} \zeta_5 \Bigg\}  
                                            \nonumber\\
                                          &+
                                            {\dis{C_{A} C_{F}^2}} \Bigg\{ \frac{352}{\epsilon^5} + 
                                            \Bigg( - \frac{2888}{9} - 32
                                            \zeta_2 \Bigg)
                                            \frac{1}{\epsilon^4} + \Bigg(
                                            \frac{4436}{27}  - 108 \zeta_2 +
                                            208 \zeta_3 \Bigg)
                                            \frac{1}{\epsilon^3} 
                                            \nonumber\\
                                          &+
                                            \Bigg( \frac{39844}{81}  +
                                            \frac{983}{9} \zeta_2  -
                                            \frac{332}{5} \zeta_2^2  - 
                                            \frac{1928}{9} \zeta_3 \Bigg)
                                            \frac{1}{\epsilon^2}  + \Bigg( -
                                            \frac{97048}{243}  -
                                            \frac{12361}{54} \zeta_2  +
                                            \frac{2975}{36} \zeta_2^2  
                                            \nonumber\\
                                          &+ 
                                            \frac{3227}{3} \zeta_3 -
                                            \frac{430}{3} \zeta_2 \zeta_3  +
                                            284 \zeta_5 \Bigg)  \frac{1}{\epsilon}   -
                                            \frac{709847}{729} +
                                            \frac{36845}{324} \zeta_2  -
                                            \frac{536683}{2160} \zeta_2^2  - 
                                            \frac{18619}{1260} \zeta_2^3  
                                            \nonumber\\
                                          &-
                                            \frac{31537}{18} \zeta_3  -
                                            \frac{518}{3} \zeta_2 \zeta_3  + 
                                            \frac{1616}{3} \zeta_3^2  + 
                                            \frac{1750}{9} \zeta_5 \Bigg\}\,.
\end{align}

\subsection{Universal Behaviour of Leading Transcendentality
  Contribution}
\label{sec:SUYM}

In \cite{Gehrmann:2011xn}, the form factor of a scalar composite
operator belonging to the stress-energy tensor super-multiplet of
conserved currents of ${\cal N}=4$ super Yang-Mills (SYM) with gauge
group SU(N) was studied to three-loop level. Since the theory is UV
finite in $d=4$ space-time dimensions, it is an ideal framework to
study the IR structures of amplitudes in perturbation theory.  In this
theory, one observes that scattering amplitudes can be expressed as a
linear combinations of polylogarithmic functions of uniform degree
$2 l$, where $l$ is the order of the loop, with constant coefficients.
In other words, the scattering amplitudes in ${\cal N}=4$ SYM exhibit
uniform transcendentality, in contrast to QCD loop amplitudes, which
receive contributions from all degrees of transcendentality up to
$2l$.

The three-loop QCD quark and gluon form factors~\cite{Gehrmann:2010ue}
display an interesting relation to the SYM form factor. Upon
replacement~\cite{hep-th/0611204} of the color factors $C_A = C_F = N$
and $T_{f} n_{f}=N/2$, the
leading transcendental (LT) parts of the quark and gluon form factors
in QCD not only coincide with each other but also become identical, up
to a normalization factor of $2^{l}$, to the form factors of scalar
composite operator computed in ${\cal N}=4$ SYM
\cite{Gehrmann:2011xn}.

This correspondence between the QCD form factors and that of the
${\cal N}=4$ SYM can be motivated by the leading transcendentality
principle~\cite{hep-th/0611204, hep-th/0404092, Kotikov:2001sc} which
relates anomalous dimensions of the twist two operators in
${\cal N} =4$ SYM to the LT terms of such operators computed in QCD.
% However, unlike the case for ${\cal N}=1$, the quark and gluon form
% factors in QCD get additional contributions arising from diagrams
% with scalar particles in ${\cal N}=2$ and ${\cal N}=4$
% SYM~\cite{Kotikov:2001sc}.
Examining the diagonal pseudo-scalar form factors ${\cal F}^G_g$ and
${\cal F}^J_q$, we find a similar behaviour: the LT terms of these
form factors with replacement $C_A = C_F = N$ and 
  $T_{f} n_{f}=N/2$ are not only identical to
each other but also coincide with the LT terms of the QCD form
factors~\cite{Gehrmann:2010ue} with the same replacement as well as
with the LT terms of the scalar form factors in ${\cal N}=4$ SYM
\cite{Gehrmann:2011xn}, up to a normalization factor of
  $2^{l}$. This observation holds true for the finite terms in
$\epsilon$, and could equally be validated for higher-order terms up
to transcendentality 8 (which is the highest order for which all
three-loop master integrals are available~\cite{Lee:2010ik}).  In
addition to checking the diagonal form factors, we also examined the
off-diagonal ones namely, ${\cal F}^{G}_{q}$, ${\cal F}^{J}_{g}$,
where we find that the LT terms these two form factors are identical
to each other after the replacement of colour factors. However, the LT
terms of these do not coincide with those of the diagonal ones.

\section{Hard Matching Coefficients in SCET}
\label{sec:scet}

Soft-collinear effective theory (SCET, \cite{hep-ph/0005275,
  hep-ph/0011336, hep-ph/0107001, hep-ph/0109045, hep-ph/0206152,
  hep-ph/0211358, hep-ph/0202088}) is a systematic expansion of the
full QCD theory in terms of particle modes with different infrared
scaling behaviour. It provides a framework to perform threshold
resummation. In the effective theory, the infrared poles of the full
high energy QCD theory manifest themselves as ultraviolet
poles~\cite{Korchemsky:1985xj, Korchemsky:1987wg, Korchemsky:1988pn},
which then can be resummed by employing the renormalisation group
evolution from larger scales to the smaller ones. To ensure matching
of SCET and full QCD, one computes the matrix elements in both
theories and adjusts the Wilson coefficients of SCET accordingly. For
the on-shell matching of these two theories, the matching coefficients
relevant to pseudo-scalar production in gluon fusion can be obtained
directly from the gluon form factors.

The UV renormalised form factors in QCD contain infrared (IR)
divergences. Since the IR poles in QCD turn into UV ones in SCET, we
can remove the IR divergences with the help of a renormalisation
constant $Z^{A,h}_{g}$, which essentially absorbs all residual IR poles
and produces finite results. The result is the matching coefficient
$C^{A, {\rm eff}}_{g}$, which is defined through the following
factorisation relation:
\begin{align}
  \label{eq:MatchCof}
  C^{A, {\rm eff}}_{g}\l(Q^{2},\mu_h^{2}\r) &\equiv \lim_{\epsilon
                                              \rightarrow 0} (Z_{g
                                              }^{A,h})^{-1}(\epsilon, Q^{2},\mu_h^{2})
                                              \left[ {\cal F}^{A}_{g} \right]_{R}(\epsilon,Q^{2})
\end{align}  
where, the UV renormalised form factor
$\left[ {\cal F}^{A}_{g} \right]_{R}$, is defined as
\begin{align}
  \label{eq:UVRenFF}
  \left[ {\cal F}^{A}_{g} \right]_{R} =   \left[ {\cal F}^{G}_{g}
  \right]_{R}  + \frac{4 C_J}{C_G}  \left[ {\cal F}^{J}_{g}
  \right]_{R} { \l( a_{s} \frac{S^{J,(1)}_g}{S^{G,(0)}_{g}}
  \r)} \,.
\end{align}
The parameter $\mu_h$ is the newly introduced mass scale at which the
above factorisation is carried out. For the UV renormalised form
factors $[{\cal F}^{A}_{g}]_{R}$ in Eq.~(\ref{eq:MatchCof}), we fixed
the other scales as $\mu_{R}^{2}=\mu_{F}^{2}=\mu_h^2$. Upon
expanding the $Z^{A,h}_{g}$ and $C^{A,{\rm eff}}_{g}$ in powers
of $a_{s}$ as
\begin{align}
  \label{eq:MatchCofExp}
  Z^{A,h}_{g}(\epsilon, Q^{2},\mu_h^{2}) &= 1 + \sum_{i=1}^{\infty}
                                                  a_{s}^{i}(\mu_h^{2})
                                                  Z^{A, h}_{g,i}(\epsilon,
                                                  Q^{2}, \mu_h^{2})\,,
                                                  \nonumber\\
  C^{A, {\rm eff}}_{g}\l(Q^{2},\mu_h^{2}\r) &= 1+\sum_{i=1}^{\infty}
                                              a_{s}^{i}(\mu_h^{2})
                                              C^{A, {\rm eff}}_{g,i}\l(Q^{2},\mu_h^{2}\r)
\end{align}
and utilising the above Eq.~(\ref{eq:MatchCof}), we compute the
$Z^{A,h}_{g,i}$ as well as $C^{A, {\rm eff}}_{g,i}$ up to three
loops ($i=3$). Demanding the cancellation of the residual IR poles of
$\left[ {\cal F}^{A}_{g} \right]_{R}$ against the poles of
$(Z^{A,h}_{g,i})^{-1}$, we compute $Z^{A,h}_{g,i}$ which
comes out to be
\begin{align}
  \label{eq:ZIR}
  Z^{A,h}_{g,1} &= {\dis{C_{A}}} \Bigg\{ - \frac{8}{\epsilon^2} +
                         \Bigg( - 4 L +
                         \frac{22}{3}   \Bigg) \frac{1}{\epsilon}
                         \Bigg\}  - {\dis{n_{f}}} \Bigg\{ \frac{4}{3 \epsilon} \Bigg\}\,,
                         \nonumber\\
  Z^{A,h}_{g,2} &= C_{F} n_{f} \Bigg\{ - \frac{2}{\epsilon} \Bigg\}
                         +  n_{f}^2 \Bigg\{ \frac{16}{9
                         \epsilon^2}  \Bigg\}
                         + 
                         C_{A} n_{f} \Bigg\{ \frac{56}{3 \epsilon^3} +
                         \Bigg( - \frac{52}{3} +  8 L \Bigg) \frac{1}{\epsilon^2} + 
                         \Bigg( - \frac{128}{27} + \frac{20}{9} L 
                         \nonumber\\
                       &+
                         \frac{2}{3} \zeta_2 \Bigg)
                         \frac{1}{\epsilon} \Bigg\} 
                         + 
                         C_{A}^2 \Bigg\{ \frac{32}{\epsilon^4} + \Bigg(
                         - \frac{308}{3} + 32 L \Bigg)  \frac{1}{\epsilon^3} + 
                         \Bigg( \frac{350}{9} - 44 L + 8 L^2 + 4
                         \zeta_2 \Bigg) \frac{1}{\epsilon^2}  + 
                         \Bigg( \frac{692}{27} 
                         \nonumber\\
                       &- \frac{134}{9} L -
                         \frac{11}{3} \zeta_2 +  4 L \zeta_2 - 2
                         \zeta_3 \Bigg)  \frac{1}{\epsilon} \Bigg\}\,,
                         \nonumber\\
  Z^{A,h}_{g,3} &= C_{F}^2 n_{f} \Bigg\{ \frac{2}{3 \epsilon}
                         \Bigg\} 
                         + C_{F} n_{f}^2
                         \Bigg\{ \frac{56}{9 \epsilon^2}  +
                         \frac{22}{27 \epsilon} \Bigg\}   
                         - 
                         n_{f}^3 \Bigg\{ \frac{64}{27 \epsilon^3}
                         \Bigg\} 
                         + C_{A}^2
                         n_{f} \Bigg\{  - \frac{320}{3 \epsilon^5} +
                         \Bigg( \frac{28480}{81} 
                         \nonumber\\
                       &-  96 L \Bigg) \frac{1}{\epsilon^4} + 
                         \Bigg( - \frac{18752}{243} + \frac{3152}{27}
                         L - \frac{64}{3} L^2  - \frac{448}{27}
                         \zeta_2 \Bigg) \frac{1}{\epsilon^3}  + 
                         \Bigg( - \frac{32656}{243} + \frac{7136}{81}
                         L 
                         \nonumber\\
                       &- \frac{80}{9} L^2  + \frac{1000}{81} \zeta_2 - 
                         \frac{104}{9} L \zeta_2 + \frac{344}{27}
                         \zeta_3 \Bigg) \frac{1}{\epsilon^2}  + 
                         \Bigg( - \frac{30715}{2187} + \frac{836}{81}
                         L + \frac{2396}{243} \zeta_2  -
                         \frac{160}{27} L \zeta_2 
                         \nonumber\\
                       &-  
                         \frac{328}{45} \zeta_2^2 - \frac{712}{81}
                         \zeta_3  + \frac{112}{9} L \zeta_3 \Bigg)
                         \frac{1}{\epsilon} \Bigg\} 
                         +  
                         C_{A} n_{f}^2 \Bigg\{ - \frac{2720}{81
                         \epsilon^4}  + \Bigg( \frac{7984}{243} -
                         \frac{352}{27} L \Bigg)  \frac{1}{\epsilon^3} + 
                         \Bigg( \frac{368}{27} 
                         \nonumber\\
                       &- \frac{400}{81} L -
                         \frac{40}{27} \zeta_2  \Bigg)
                         \frac{1}{\epsilon^2} +  
                         \Bigg( \frac{269}{2187} + \frac{16}{81} L -
                         \frac{40}{81} \zeta_2  + \frac{112}{81}
                         \zeta_3 \Bigg) \frac{1}{\epsilon}  \Bigg\} 
                         + C_{A} 
                         C_{F} n_{f} \Bigg\{ \frac{272}{9 \epsilon^3} 
                         \nonumber\\
                       &+
                         \Bigg( - \frac{704}{27}  + \frac{40}{3} L -
                         \frac{64}{9} \zeta_3 \Bigg)
                         \frac{1}{\epsilon^2} +  
                         \Bigg( - \frac{2434}{81} + \frac{110}{9} L +
                         \frac{4}{3} \zeta_2  + \frac{32}{15}
                         \zeta_2^2 +  
                         \frac{304}{27} \zeta_3 
                         \nonumber\\
                       &- \frac{32}{3} L
                         \zeta_3 \Bigg)  \frac{1}{\epsilon} \Bigg\}
                         +  
                         C_{A}^3 \Bigg\{ - \frac{256}{3 \epsilon^6} +
                         \Bigg( \frac{1760}{3}  - 128 L \Bigg)
                         \frac{1}{\epsilon^5} +  
                         \Bigg( - \frac{72632}{81} + 528 L - 64 L^2
                         \nonumber\\
                       &- 32 \zeta_2 \Bigg)  \frac{1}{\epsilon^4} + 
                         \Bigg( - \frac{29588}{243} - \frac{5824}{27}
                         L + \frac{352}{3} L^2  - \frac{32}{3} L^3 + 
                         \frac{2464}{27} \zeta_2 - 48 L \zeta_2 + 16
                         \zeta_3 \Bigg)  \frac{1}{\epsilon^3} 
                         \nonumber\\
                       &+ 
                         \Bigg( \frac{80764}{243} - \frac{25492}{81}
                         L + \frac{536}{9} L^2  - \frac{1486}{81}
                         \zeta_2 +  
                         \frac{572}{9} L \zeta_2 - 16 L^2 \zeta_2 -
                         \frac{352}{45} \zeta_2^2  - \frac{836}{27}
                         \zeta_3 
                         \nonumber\\
                       &+  
                         8 L \zeta_3 \Bigg) \frac{1}{\epsilon^2} +
                         \Bigg( \frac{194372}{2187}  - \frac{490}{9}
                         L - \frac{12218}{243} \zeta_2  + 
                         \frac{1072}{27} L \zeta_2 + \frac{1276}{45}
                         \zeta_2^2  - \frac{176}{15} L \zeta_2^2 - 
                         \frac{244}{9} \zeta_3 
                         \nonumber\\
                       &- \frac{88}{9} L
                         \zeta_3  + \frac{80}{9} \zeta_2 \zeta_3  +
                         \frac{32}{3} \zeta_5 \Bigg)  \frac{1}{\epsilon} \Bigg\}\,.
\end{align}
After cancellation of the IR poles, we are left with the following
finite matching coefficients:
\begin{align}
  \label{eq:SCETCof}
  C^{A, {\rm eff}}_{g,1} &= {\dis{C_{A}}} \Bigg\{ - L^2 + 4 + \zeta_2 \Bigg\}\,,
                           \nonumber\\
  C^{A, {\rm eff}}_{g,2} &=  C_{A}^2 \Bigg\{  \frac{1}{2} L^4 + 
                           \frac{11}{9} L^3  + L^2
                           \Bigg( - \frac{103}{9}  + \zeta_2 \Bigg) 
                           + L \Bigg( - \frac{10}{27}  - \frac{22}{3} 
                           \zeta_2 - 2 \zeta_3 \Bigg) + \frac{4807}{81} + 
                           \frac{91}{6} \zeta_2 
                           \nonumber\\
                         &+ \frac{1}{2}\zeta_2^2 -  
                           \frac{143}{9} \zeta_3 \Bigg\} 
                           + C_{A} n_{f}
                           \Bigg\{  - \frac{2}{9} L^3 +  \frac{10}{9}
                           L^2 + L \Bigg(
                           \frac{34}{27} + \frac{4}{3} \zeta_2 \Bigg)
                           - \frac{943}{81} -  
                           \frac{5}{3} \zeta_2 - \frac{46}{9} \zeta_3 \Bigg\}  
                           \nonumber\\
                         & + 
                           C_{F} n_{f} \Bigg\{ - \frac{80}{3} + 6
                           \ln \left(\frac{\mu_h^2}{m_t^2}\right) + 8 \zeta_3 \Bigg\}\,,
                           \nonumber\\
  C^{A, {\rm eff}}_{g,3} &= n_{f} C^{(2)}_{J} \Bigg\{ - {2}
                           \Bigg\} 
                           + C_{F} n_{f}^2 \Bigg\{ L \Bigg( -
                           \frac{320}{9} + 8 \ln \left(\frac{\mu_h^2}{m_t^2}\right) +  \frac{32}{3}
                           \zeta_3 \Bigg) +
                           \frac{749}{9} - \frac{20}{9} \zeta_2  -
                           \frac{16}{45} \zeta_2^2  
                           \nonumber\\
                         &- 
                           \frac{112}{3} \zeta_3  \Bigg\}  
                           + 
                           C_{F}^2 n_{f} \Bigg\{ \frac{457}{6} + 104
                           \zeta_3 - 160 \zeta_5 \Bigg\}  
                           + 
                           C_{A}^2 n_{f} \Bigg\{  
                           \frac{2}{9} L^5 - \frac{8}{27} L^4 + 
                           L^3 \Bigg( - \frac{752}{81} 
                           \nonumber\\
                         &- \frac{2}{3}
                           \zeta_2 \Bigg)  + 
                           L^2 \Bigg( \frac{512}{27} - \frac{103}{9}
                           \zeta_2  + \frac{118}{9} \zeta_3 \Bigg) +  
                           L \Bigg( \frac{129283}{729} +
                           \frac{4198}{81} \zeta_2 -  \frac{48}{5}
                           \zeta_2^2 +  \frac{28}{9} \zeta_3 \Bigg) 
                           \nonumber\\
                         &-
                           \frac{7946273}{13122} -  \frac{19292}{729} \zeta_2
                           + \frac{73}{45} \zeta_2^2  - 
                           \frac{2764}{81} \zeta_3 - \frac{82}{9}
                           \zeta_2 \zeta_3  +
                           \frac{428}{9} \zeta_5 \Bigg\} 
                           +  
                           C_{A}^3 \Bigg\{  - \frac{1}{6} L^6 -
                           \frac{11}{9} L^5   
                           \nonumber\\
                         &+ L^4
                           \Bigg( \frac{389}{54}  - \frac{3}{2}
                           \zeta_2 \Bigg) + 
                           L^3 \Bigg( \frac{2206}{81} + \frac{11}{3}
                           \zeta_2 + 2 \zeta_3 \Bigg) +  
                           L^2 \Bigg( - \frac{20833}{162} +
                           \frac{757}{18} \zeta_2  - \frac{73}{10}
                           \zeta_2^2  
                           \nonumber\\
                         &+ \frac{143}{9} \zeta_3 \Bigg) + 
                           \frac{2222}{9} \zeta_5 + L \Bigg( -
                           \frac{500011}{1458}  - \frac{16066}{81}
                           \zeta_2 + \frac{176}{5} \zeta_2^2  + 
                           \frac{1832}{27} \zeta_3 + \frac{34}{3}
                           \zeta_2 \zeta_3  
                           \nonumber\\
                         &+ 16 \zeta_5 \Bigg) + \frac{41091539}{26244} + 
                           \frac{316939}{1458} \zeta_2 -
                           \frac{1399}{270} \zeta_2^2 -
                           \frac{24389}{1890} \zeta_2^3 -  
                           \frac{176584}{243} \zeta_3 - \frac{605}{9}
                           \zeta_2 \zeta_3  
                           \nonumber\\
                         &- \frac{104}{9} \zeta_3^2  \Bigg\} 
                           + 
                           C_{A} n_{f}^2 \Bigg\{
                           - \frac{2}{27} L^4 + \frac{40}{81} L^3   + 
                           L^2 \Bigg( \frac{80}{81} + \frac{8}{9}
                           \zeta_2 \Bigg) + 
                           L \Bigg( - \frac{12248}{729} -
                           \frac{80}{27} \zeta_2 
                           \nonumber\\
                         &- \frac{128}{27}
                           \zeta_3 \Bigg) + \frac{280145}{6561} +
                           \frac{4}{9}  \zeta_2  + \frac{4}{27} \zeta_2^2  +  \frac{4576}{243} \zeta_3
                           \Bigg\}
                           + 
                           C_{A} C_{F} n_{f} \Bigg\{  -
                           \frac{2}{3} L^3  + L^2 \Bigg( \frac{215}{6}
                           \nonumber\\
                         &- 6 \ln \left(\frac{\mu_h^2}{m_t^2}\right)
                           - 16 \zeta_3 \Bigg) +
                           L \Bigg( \frac{9173}{54}  - 44 \ln
                           \left(\frac{\mu_h^2}{m_t^2}\right)  + 4 \zeta_2 + 
                           \frac{16}{5} \zeta_2^2 - \frac{376}{9}
                           \zeta_3 \Bigg) 
                           \nonumber\\
                         &+ 24 \ln
                           \left(\frac{\mu_h^2}{m_t^2}\right)  -
                           \frac{726935}{972} -
                           \frac{415}{18} \zeta_2  + 
                           6 \ln \left(\frac{\mu_h^2}{m_t^2}\right)
                           \zeta_2 - \frac{64}{45} \zeta_2^2    +  
                           \frac{20180}{81} \zeta_3 + \frac{64}{3}
                           \zeta_2 \zeta_3 
                           \nonumber\\
                         &+  \frac{608}{9} \zeta_5 \Bigg\}\,. 
\end{align}
In the above expressions,
$L=\ln \left( Q^{2}/\mu_h^{2} \right)=\ln \left( -q^{2}/\mu_h^{2}
\right)$.
These matching coefficients allow to perform the matching of the
SCET-based resummation onto the full QCD calculation up to three-loop
order.

Before ending the discussion of this section, we demonstrate the
universal factorisation property fulfilled by the anomalous dimension
of the $Z^{A,h}_{g}$ which is defined through the RG equation
\begin{align}
  \label{eq:RGEZIR}
  \mu_{h}^{2}\frac{d}{d\mu_{h}^{2}} \ln Z^{A,h}_{g}(\epsilon, Q^{2},\mu_{h}^{2}) \equiv
  \gamma^{A,h}_{g}(Q^{2},\mu_{h}^{2}) = \sum_{i=1}^{\infty} a_{s}^{i}(\mu_{h}^{2})
  \gamma^{A,h}_{g,i}(Q^{2},\mu_{h}^{2})\,.
\end{align}
The renormalisation group invariance of the UV renormalised
$[F^{A}_{g}]_{R}(\epsilon, Q^{2})$ with respect to the scale $\mu_{h}$
implies
\begin{align}
  \label{eq:RGEZIRC}
  \mu_{h}^{2}\frac{d}{d\mu_{h}^{2}} \ln Z^{A,h}_{g} +
  \mu_{h}^{2}\frac{d}{d\mu_{h}^{2}} \ln C^{A,{\rm eff}}_{g} = 0\,. 
\end{align}  
By explicitly evaluating the $\gamma^{A,h}_{g,i}$ using the
results of $Z^{A,h}_{g}$ (Eq.~(\ref{eq:ZIR})) up to three loops
($i=3$), we find that these satisfy the following decomposition in
terms of the universal factors $A_{g,i}, B_{g,i}$ and $f_{g.i}$:
\begin{align}
  \label{eq:gammaIRdecom}
  \gamma^{A,h}_{g,i} = - \frac{1}{2} A_{g,i} L + \l(B_{g,i} +
  \frac{1}{2} f_{g,i}\r)\,.
\end{align} 
This in turn implies the evolution equation of the matching
coefficients as
\begin{align}
  \label{eq:RGECeff}
  \mu_{h}^{2}\frac{d}{d\mu_{h}^{2}} \ln C^{A,{\rm eff}}_{g,i} = \frac{1}{2} A_{g,i} L - \l(B_{g,i} +
  \frac{1}{2} f_{g,i}\r)
\end{align}
which is in complete agreement with the existing
results~\cite{Becher:2006nr} upon identifying
\begin{align}
  \label{eq:gammaV}
  \gamma^{V} = B_{g.i} + \frac{1}{2} f_{g,i}\,.
\end{align}

\section{Conclusions}
\label{sec:conc}
In this paper, we derived the three-loop massless QCD corrections to
the quark and gluon form factors of pseudo-scalar operators.  Working
in dimensional regularisation, we used the 't~Hooft-Veltman
prescription for $\gamma_5$ and the Levi-Civita tensor, which requires
non-trivial finite renormalisation to maintain the symmetries of the
theory. By exploiting the universal behaviour of the infrared pole
structure at three loops in QCD, we were able to independently
determine the renormalisation constants and operator mixing, in
agreement with earlier results that were obtained in a completely
different approach~\cite{Larin:1993tq,Zoller:2013ixa}.

The three-loop corrections to the pseudo-scalar form factors are an
important ingredient to precision Higgs phenomenology. They will
ultimately allow to bring the gluon fusion cross section for
pseudo-scalar Higgs production to the same level of accuracy that has
been accomplished most recently for scalar Higgs production with fixed
order N$^3$LO~\cite{Anastasiou:2015ema} and soft-gluon resummation at
N$^3$LL~\cite{Ahrens:2008nc,Bonvini:2014joa,Catani:2014uta,Schmidt:2015cea}.

With our new results, the soft-gluon resummation for pseudo-scalar
Higgs production~\cite{deFlorian:2007sr,Schmidt:2015cea} can be
extended imminently to N$^3$LL
{accuracy~\cite{Ahmed:2015pSSV}}, given the established
formalisms at this order~\cite{Ahrens:2008nc,Catani:2014uta}. 
With the derivation of the three-loop pseudo-scalar form factors 
presented here, all ingredients to this calculation are now available. 
Another imminent application is
the threshold approximation to the N$^3$LO cross {section~\cite{Ahmed:2015pSSV}}.  By
exploiting the universal infrared structure~\cite{Catani:2014uta}, one
can use the result of an explicit computation of the threshold
contribution to the N$^3$LO cross section for scalar Higgs
production~\cite{Anastasiou:2014vaa} to derive threshold results for
other processes essentially through the ratios of the respective form
factors (which is no longer possible beyond
threshold~\cite{Anastasiou:2014lda,Anastasiou:2015ema}, where the
corrections become process-specific), as was done for the Drell-Yan
process~\cite{Ahmed:2014cla} and for Higgs production from bottom
quark annihilation~\cite{Ahmed:2014cha}. These will be investigated in
a separate publication.

\section*{Acknowledgments}
We are grateful to Roman N. Lee for useful discussions and timely help with 
the LiteRed package.  
We would like to thank Michael Spira, Massimiliano Grazzini and 
M.~C. Kumar for
useful discussions and comments on the manuscript. This research was
supported in part by the Swiss National Science Foundation (SNF) under
contract 200020-162487, as well as by the European Commission through
the ERC Advanced Grant ``MC@NNLO" (340983). 
\appendix

\section{Results of Unrenormalised Form Factors}
\label{app:Results}

In this part, we present the unrenormalised quark and gluon form
factors up to three loops for the operators $\left[ O_{G} \right]_{B}$
and $\left[ O_{J} \right]_{B}$. Specifically, we present
${\hat{\cal F}}^{G,(n)}_{\beta}$ and ${\hat{\cal F}}^{J,(n)}_{\beta}$
for $\beta=q,g$ up to $n=3$ which are defined in
Sec.~\ref{sec:FF}. One and two loop results completely agree with the
existing literature~\cite{Ravindran:2004mb}. It should be noted
that the form factors at $n=2$ for ${\hat{\cal F}}^{G,(n)}_{q}$ and
${\hat{\cal F}}^{J,(n)}_{g}$ correspond to the contributions arising
from three loop diagrams since these processes start at one loop order.

% ***** FgG1 *****
%
\begin{align}
  \label{eq:FgG1}
  {\hat{\cal F}}^{G,(1)}_{g} &= {\dis{C_{A}}} \Bigg\{ - \frac{8}{\epsilon^2} + 4 + \zeta_2 + 
                               \epsilon \Bigg( - 6 - \frac{7}{3} \zeta_3 \Bigg) + \epsilon^2 \Bigg(
                               7 - \frac{\zeta_2}{2} + \frac{47}{80}
                               \zeta_2^2 \Bigg) + \epsilon^3 \Bigg( -
                               \frac{15}{2} + \frac{3}{4} \zeta_2 
                               \nonumber\\
                             &+ \frac{7}{6} \zeta_3 + 
                               \frac{7}{24} \zeta_2 \zeta_3 - \frac{31}{20} \zeta_5 \Bigg) \Bigg\}\,,
\end{align}

% **** FgG2 ****
%
\begin{align}
  \label{eq:FgG2}
  {\hat{\cal F}}^{G,(2)}_{g} &= {\dis{2 C_{A} n_{f} T_{F}}} \Bigg\{ - \frac{8}{3 \epsilon^3} + \frac{20}{9 \epsilon^2} + 
                               \Bigg( \frac{106}{27} + 2 \zeta_2
                               \Bigg) \frac{1}{\epsilon}  -
                               \frac{1591}{81}  - \frac{5}{3} \zeta_2
                               - \frac{74}{9} \zeta_3 + 
                               \epsilon \Bigg( \frac{24107}{486}  
\nonumber\\
&-
                               \frac{23}{18} \zeta_2  
                               + \frac{51}{20} \zeta_2^2 + \frac{383}{27} \zeta_3 \Bigg) + 
                               \epsilon^2 \Bigg( - \frac{146147}{1458}
                               + \frac{799}{108} \zeta_2 - \frac{329}{72} \zeta_2^2 - 
                               \frac{1436}{81} \zeta_3  + \frac{25}{6}
                               \zeta_2 \zeta_3  
                               \nonumber\\
                             &- \frac{271}{30} \zeta_5 \Bigg) \Bigg\} 
                               + 
                               {\dis{C_{A}^2}} \Bigg\{ \frac{32}{\epsilon^4}
                               + \frac{44}{3 \epsilon^3}  + \Bigg( -
                               \frac{422}{9}  - 4 \zeta_2 \Bigg) \frac{1}{\epsilon^2} + 
                               \Bigg( \frac{890}{27} - 11 \zeta_2  +
                               \frac{50}{3} \zeta_3 \Bigg)
                               \frac{1}{\epsilon} 
                               \nonumber\\
                             &+ \frac{3835}{81} + 
                               \frac{115}{6} \zeta_2 - \frac{21}{5}
                               \zeta_2^2  + \frac{11}{9} \zeta_3 + 
                               \epsilon \Bigg( - \frac{213817}{972} -
                               \frac{103}{18} \zeta_2  +
                               \frac{77}{120} \zeta_2^2  +
                               \frac{1103}{54} \zeta_3  
                               \nonumber\\
                             &- 
                               \frac{23}{6} \zeta_2 \zeta_3  -
                               \frac{71}{10} \zeta_5 \Bigg)  + 
                               \epsilon^2 \Bigg( \frac{6102745}{11664}
                               - \frac{991}{27} \zeta_2  -
                               \frac{2183}{240} \zeta_2^2  + 
                               \frac{2313}{280} \zeta_2^3  -
                               \frac{8836}{81} \zeta_3   -
                               \frac{55}{12} \zeta_2 \zeta_3  
                               \nonumber\\
                             &+ 
                               \frac{901}{36} \zeta_3^2  + \frac{341}{60} \zeta_5 \Bigg) \Bigg\} 
                               + 
                               {\dis{2 C_{F} n_{f}  T_{F}}} \Bigg\{ \frac{12}{\epsilon}
                               - \frac{125}{3} + 8 \zeta_3  + \epsilon
                               \Bigg( \frac{3421}{36}  - \frac{14}{3}
                               \zeta_2  - \frac{8}{3} \zeta_2^2 
\nonumber\\
&- 
                               \frac{64}{3} \zeta_3 \Bigg) 
                              + \epsilon^2 \Bigg( - \frac{78029}{432}
                               + \frac{293}{18} \zeta_2 + 
                               \frac{64}{9} \zeta_2^2  +
                               \frac{973}{18} \zeta_3  - \frac{10}{3}
                               \zeta_2 \zeta_3  + 8 \zeta_5 \Bigg) \Bigg\}\,,
\end{align}
%

% *** FgG3 ***

\begin{align}
  \label{eq:FgG3}
  {\hat{\cal F}}^{G,(3)}_{g} &= {\dis{4 C_{F} n_{f}^2 T_{F}^{2}}} \Bigg\{ 
                               \frac{16}{\epsilon^2} + \Bigg( -
                               \frac{796}{9} + \frac{64}{3} \zeta_3
                               \Bigg) \frac{1}{\epsilon} +
                               \frac{8387}{27} - \frac{38}{3} \zeta_2
                               - \frac{112}{15} \zeta_2^2 - 
                               \frac{848}{9} \zeta_3  \Bigg\} 
                               \nonumber\\
&+ {\dis{2 C_{F}^2 n_{f} T_{F}}} \Bigg\{ 
                               \frac{6}{\epsilon} 
                              - \frac{353}{6} + 176 \zeta_3 - 160 \zeta_5 \Bigg\} + 
                               {\dis{2 C_{A}^2 n_{f}  T_{F}}} \Bigg\{ \frac{64}{3
                               \epsilon^5} - \frac{32}{81 \epsilon^4} 
                               + \Bigg( - \frac{18752}{243}  
                               \nonumber\\
&- \frac{376}{27} \zeta_2 \Bigg) \frac{1}{\epsilon^3} + 
                               \Bigg( \frac{36416}{243} 
                              -
                               \frac{1700}{81} \zeta_2 +
                               \frac{2072}{27} \zeta_3 \Bigg)
                               \frac{1}{\epsilon^2} + \Bigg(
                               \frac{62642}{2187} + \frac{22088}{243} \zeta_2 - \frac{2453}{90}
                               \zeta_2^2 
\nonumber\\
&-
                               \frac{3988}{81} \zeta_3 \Bigg)
                               \frac{1}{\epsilon} -
                               \frac{14655809}{13122} 
                              - \frac{60548}{729} \zeta_2 + 
                               \frac{917}{60} \zeta_2^2  -
                               \frac{772}{27} \zeta_3 - \frac{439}{9}
                               \zeta_2 \zeta_3  
                               + \frac{3238}{45}
                               \zeta_5 \Bigg\}  
                               \nonumber\\
&+ 
                               {\dis{4 C_{A} n_{f}^2 T_{F}^{2}}} \Bigg\{ - \frac{128}{81
                               \epsilon^4} + \frac{640}{243
                               \epsilon^3} 
                              + \Bigg(
                               \frac{128}{27} + \frac{80}{27} \zeta_2
                               \Bigg) \frac{1}{\epsilon^2} + 
                               \Bigg( - \frac{93088}{2187} -
                               \frac{400}{81} \zeta_2 
\nonumber\\
&-
                               \frac{1328}{81} \zeta_3 \Bigg)
                               \frac{1}{\epsilon} +
                               \frac{1066349}{6561} - \frac{56}{27} \zeta_2 
                              + 
                               \frac{797}{135} \zeta_2^2  
                               +\frac{13768}{243} \zeta_3 \Bigg\} 
                               +
                               {\dis{2 C_{A} C_{F} n_{f} T_{F}}} \Bigg\{ -
                               \frac{880}{9 \epsilon^3} 
\nonumber\\
&+ \Bigg(
                               \frac{6844}{27} - \frac{640}{9} \zeta_3
                               \Bigg) \frac{1}{\epsilon^2} + \Bigg( -
                               \frac{16219}{81} 
                              + \frac{158}{3} \zeta_2 
                               + \frac{352}{15} \zeta_2^2 +
                               \frac{1744}{27} \zeta_3 \Bigg)
                               \frac{1}{\epsilon} - \frac{753917}{972}  
                               \nonumber\\
&- \frac{593}{6} \zeta_2 - \frac{96}{5} \zeta_2^2  + 
                               \frac{4934}{81} \zeta_3 + 48 \zeta_2
                               \zeta_3  
                              + \frac{32}{9} \zeta_5 \Bigg\} 
                               + 
                               {\dis{C_{A}^3}} \Bigg\{ - \frac{256}{3
                               \epsilon^6} - \frac{352}{3 \epsilon^5}
                               + \frac{16144}{81 \epsilon^4} 
\nonumber\\
&+ 
                               \Bigg( \frac{22864}{243} +
                               \frac{2068}{27} \zeta_2 - \frac{176}{3}
                               \zeta_3 \Bigg) \frac{1}{\epsilon^3} 
                              + 
                               \Bigg( - \frac{172844}{243} -
                               \frac{1630}{81} \zeta_2 +
                               \frac{494}{45} \zeta_2^2 -
                               \frac{836}{27} \zeta_3 \Bigg)
                               \frac{1}{\epsilon^2} 
                               \nonumber\\
&+ \Bigg( \frac{2327399}{2187} -
                               \frac{71438}{243} \zeta_2 
                              + 
                               \frac{3751}{180} \zeta_2^2 -
                               \frac{842}{9} \zeta_3 
                               + \frac{170}{9} \zeta_2 \zeta_3 + 
                               \frac{1756}{15} \zeta_5 \Bigg)
                               \frac{1}{\epsilon} 
                               + \frac{16531853}{26244} 
\nonumber\\
&+ 
                               \frac{918931}{1458} \zeta_2 +
                               \frac{27251}{1080} \zeta_2^2 
                              - \frac{22523}{270} \zeta_2^3   -
                               \frac{51580}{243} \zeta_3  +
                               \frac{77}{18} \zeta_2 \zeta_3  
                               - \frac{1766}{9} \zeta_3^2 + 
                               \frac{20911}{45} \zeta_5  \Bigg\} \,,
\end{align}

% **** FgJ2 ****

\begin{align}
  \label{eq:FgJ1}
  {\hat{\cal F}}^{J,(1)}_{g} &= {\dis{C_{A}}} \Bigg\{ - \frac{8}{\epsilon^2} +
                               4  + \zeta_2   + 
                               \epsilon \Bigg( - \frac{15}{2}  +
                               \zeta_2  - \frac{16}{3} \zeta_3 \Bigg)
                               + \epsilon^2 \Bigg( \frac{287}{24} - 2
                               \zeta_2  + \frac{127}{80} \zeta_2^2
                               \Bigg)  
                               \nonumber\\
                             &+ 
                               \epsilon^3 \Bigg( - \frac{5239}{288}  +
                               \frac{151}{48} \zeta_2  +
                               \frac{19}{120} \zeta_2^2  +
                               \frac{\zeta_3}{12}  + 
                               \frac{7}{6} \zeta_2 \zeta_3  -
                               \frac{91}{20} \zeta_5 \Bigg) \Bigg\}  + 
                               {\dis{C_{F}}} \Bigg\{ 4  + 
                               \epsilon \Bigg( - \frac{21}{2}  
                               \nonumber\\
                             &+ 6
                               \zeta_3 \Bigg)  + \epsilon^2 \Bigg(
                               \frac{155}{8}  - \frac{5}{2} \zeta_2  -
                               \frac{9}{5} \zeta_2^2  - \frac{9}{2}
                               \zeta_3 \Bigg)  + \epsilon^3 \Bigg( -
                               \frac{1025}{32}  + \frac{83}{16}
                               \zeta_2  + 
                               \frac{27}{20} \zeta_2^2  + \frac{20}{3}
                               \zeta_3  
                               \nonumber\\
                             &- \frac{3}{4} \zeta_2 \zeta_3
                               + \frac{21}{2} \zeta_5 \Bigg) \Bigg\}\,,
\end{align}

% **** FgJ2 ****

\begin{align}
  \label{eq:FgJ2}
  {\hat{\cal F}}^{J,(2)}_{g} &= {\dis{2 C_{A} n_{f} T_{F}}} \Bigg\{ - \frac{8}{3
                               \epsilon^3}  + \frac{20}{9 \epsilon^2} + 
                               \Bigg( \frac{106}{27} + 2 \zeta_2
                               \Bigg) \frac{1}{\epsilon}  -
                               \frac{1753}{81}   - \frac{\zeta_2}{3}
                               - \frac{110}{9} \zeta_3 + 
                               \epsilon \Bigg( \frac{14902}{243} 
\nonumber\\
&-
                               \frac{103}{18} \zeta_2 
                              + \frac{241}{60} \zeta_2^2 +
                               \frac{599}{27} \zeta_3 \Bigg)  + 
                               \epsilon^2 \Bigg( - \frac{411931}{2916}
                               + \frac{2045}{108} \zeta_2  -
                               \frac{2353}{360} \zeta_2^2  - 
                               \frac{3128}{81} \zeta_3 
\nonumber\\
&+ \frac{43}{6}
                               \zeta_2 \zeta_3 
                               - \frac{167}{10} \zeta_5 \Bigg) \Bigg\} + 
                               {\dis{C_{A}  C_{F}}} \Bigg\{ - \frac{32}{\epsilon^2} + 
                               \Bigg( \frac{208}{3} - 48 \zeta_3
                               \Bigg) \frac{1}{\epsilon} -
                               \frac{451}{9}  + 24 \zeta_2 +
                               \frac{72}{5} \zeta_2^2 
\nonumber\\
&- 8 \zeta_3 
                               + \epsilon \Bigg( - \frac{16385}{108} - \frac{52}{3} \zeta_2 + 
                               \frac{12}{5} \zeta_2^2 + 32 \zeta_3 +
                               10 \zeta_2 \zeta_3 - 14 \zeta_5 \Bigg) + \epsilon^2 \Bigg(
                               \frac{1073477}{1296} 
\nonumber\\
&- \frac{815}{9} \zeta_2 
                              + 
                               \frac{19}{20} \zeta_2^2  +
                               \frac{17}{70} \zeta_2^3  -
                               \frac{1915}{36} \zeta_3  + 9 \zeta_2
                               \zeta_3  - 
                               34 \zeta_3^2  - \frac{2279}{6} \zeta_5
                               \Bigg)  \Bigg\}  + 
                               {\dis{2 C_{F} n_{f} T_{F}}} \Bigg\{ \frac{26}{3
                               \epsilon} 
\nonumber\\
&- \frac{709}{18}  + 16 \zeta_3 
                              + \epsilon \Bigg( \frac{26149}{216} - \frac{65}{6} \zeta_2 - 
                               \frac{76}{15} \zeta_2^2 - 44 \zeta_3 \Bigg) + 
                               \epsilon^2 \Bigg( - \frac{828061}{2592}
                               + \frac{3229}{72} \zeta_2  
\nonumber\\
&+ \frac{212}{15} \zeta_2^2 + 
                               \frac{1729}{18} \zeta_3 
                              - 4 \zeta_2 \zeta_3 + \frac{166}{3} \zeta_5 \Bigg) \Bigg\} + 
                               {\dis{C_{A}^2}} \Bigg\{ \frac{32}{\epsilon^4}
                               + \frac{44}{3 \epsilon^3}  + \Bigg( -
                               \frac{422}{9}  - 4 \zeta_2 \Bigg)
                               \frac{1}{\epsilon^2}  
\nonumber\\
&+ 
                               \Bigg( \frac{1214}{27}  - 19 \zeta_2 
                              + \frac{122}{3} \zeta_3 \Bigg)
                               \frac{1}{\epsilon}  + \frac{1513}{81}  +
                               \frac{143}{6} \zeta_2 - \frac{61}{5}
                               \zeta_2^2  + \frac{209}{9} \zeta_3  + 
                               \epsilon \Bigg( - \frac{202747}{972}  
\nonumber\\
&+
                               \frac{59}{36} \zeta_2  - \frac{349}{24}
                               \zeta_2^2  
                              - \frac{2393}{108} \zeta_3  - 
                               \frac{53}{6} \zeta_2 \zeta_3  +
                               \frac{369}{10} \zeta_5 \Bigg)  + 
                               \epsilon^2 \Bigg( \frac{7681921}{11664}
                               - \frac{35255}{432} \zeta_2 +
                               \frac{1711}{180} \zeta_2^2  
\nonumber\\
&- 
                               \frac{7591}{840} \zeta_2^3 
                              -
                               \frac{5683}{1296} \zeta_3  -
                               \frac{407}{12} \zeta_2 \zeta_3  + 
                               \frac{775}{36} \zeta_3^2 +
                               \frac{4013}{30} \zeta_5 \Bigg) \Bigg\}
                               + 
                               {\dis{C_{F}^2}} \Bigg\{ - 6 + \epsilon \Bigg(
                               \frac{259}{12}  + 41 \zeta_3 
\nonumber\\
&- 60
                               \zeta_5 \Bigg)  
                              + 
                               \epsilon^2 \Bigg( - \frac{7697}{144}  +
                               \frac{\zeta_2}{3}  - \frac{184}{15}
                               \zeta_2^2  + \frac{120}{7} \zeta_2^3 - 
                               163 \zeta_3 + 4 \zeta_2 \zeta_3  + 30
                               \zeta_3^2  + \frac{470}{3} \zeta_5 \Bigg) \Bigg\}\,,
\end{align}

% **** FqG1 ****

\begin{align}
  \label{eq:FqG1}
  {\hat{\cal F}}^{G,(1)}_{q} &= {\dis{2 n_{f} T_{F}}} \Bigg\{ \frac{4}{3 \epsilon}  -
                               \frac{19}{9}  + \epsilon \Bigg(
                               \frac{355}{108}  - \frac{\zeta_2}{6}
                               \Bigg)  + \epsilon^2 \Bigg( -
                               \frac{6523}{1296}  + \frac{19}{72}
                               \zeta_2  + \frac{25}{18} \zeta_3 \Bigg) + 
                               \epsilon^3 \Bigg( \frac{118675}{15552}
                               \nonumber\\
                             &- \frac{355}{864} \zeta_2  -
                               \frac{191}{480} \zeta_2^2  - 
                               \frac{475}{216} \zeta_3 \Bigg) \Bigg\} 
                               + 
                               {\dis{C_{F}}} \Bigg\{ - \frac{8}{\epsilon^2}  +
                               \frac{6}{\epsilon}  - \frac{11}{2}  +
                               \zeta_2  + 
                               \epsilon \Bigg( \frac{25}{8}  -
                               \frac{3}{4} \zeta_2  - \frac{7}{3}
                               \zeta_3 \Bigg)  
                               \nonumber\\
                             &+ 
                               \epsilon^2 \Bigg( - \frac{11}{32}  -
                               \frac{21}{16} \zeta_2  + \frac{47}{80}
                               \zeta_2^2  + \frac{7}{4} \zeta_3 \Bigg) + 
                               \epsilon^3 \Bigg( - \frac{415}{128}  +
                               \frac{223}{64} \zeta_2  -
                               \frac{141}{320} \zeta_2^2  -
                               \frac{155}{48} \zeta_3  
                               \nonumber\\
                             &+ 
                               \frac{7}{24} \zeta_2 \zeta_3  -
                               \frac{31}{20}  \zeta_5 \Bigg) \Bigg\} 
                               + 
                               {\dis{C_{A}}} \Bigg\{ - \frac{22}{3 \epsilon}
                               + \frac{269}{18}   + 
                               \epsilon \Bigg( - \frac{5045}{216}  +
                               \frac{23}{12} \zeta_2  + 3 \zeta_3
                               \Bigg)  + \epsilon^2 \Bigg(
                               \frac{90893}{2592}  
                               \nonumber\\
                             &- \frac{485}{144}
                               \zeta_2  - 
                               \frac{4}{5} \zeta_2^2 - \frac{275}{36}
                               \zeta_3 \Bigg)  + 
                               \epsilon^3 \Bigg( -
                               \frac{1620341}{31104}  +
                               \frac{8861}{1728} \zeta_2  +
                               \frac{751}{320} \zeta_2^2  + 
                               \frac{4961}{432} \zeta_3 +
                               \frac{\zeta_2 \zeta_3}{8}  
                               \nonumber\\
                             &+
                               \frac{15}{2} \zeta_5  \Bigg) \Bigg\}\,,
\end{align}

% **** FqG2 ****

\begin{align}
  \label{eq:FqG2}
  {\hat{\cal F}}^{G,(2)}_{q} &= {\dis{4 n_{f}^2 T_{F}^{2}}} \Bigg\{ \frac{16}{9
                               \epsilon^2}  - \frac{152}{27 \epsilon}
                               + \frac{124}{9} - \frac{4}{9} \zeta_2  + 
                               \epsilon \Bigg( - \frac{7426}{243}  +
                               \frac{38}{27} \zeta_2  + \frac{136}{27}
                               \zeta_3 \Bigg)  + 
                               \epsilon^2 \Bigg( \frac{47108}{729}  
\nonumber\\
&-
                               \frac{31}{9} \zeta_2 
                               - \frac{43}{30}
                               \zeta_2^2  - \frac{1292}{81} \zeta_3 \Bigg)  \Bigg\} 
                               + 
                               {\dis{C_{A}^2}} \Bigg\{ \frac{484}{9
                               \epsilon^2}  - \frac{6122}{27 \epsilon}
                               + \frac{1865}{3} - \frac{319}{9}
                               \zeta_2  - 66 \zeta_3 
\nonumber\\
&+ 
                               \epsilon \Bigg( - \frac{702941}{486}  
                              +
                               \frac{14969}{108}  \zeta_2  +
                               \frac{299}{20} \zeta_2^2  + 
                               \frac{31441}{108} \zeta_3 +  5 \zeta_2
                               \zeta_3 - 30 \zeta_5 \Bigg)  + 
                               \epsilon^2 \Bigg( \frac{18199507}{5832}
                               \nonumber\\
&- \frac{5861}{16} \zeta_2 
                              -
                               \frac{63233}{720} \zeta_2^2  - 
                               \frac{691}{140} \zeta_2^3 -
                               \frac{995915}{1296} \zeta_3  +
                               \frac{52}{3} \zeta_2 \zeta_3  - 
                               \frac{39}{2} \zeta_3^2 -  \frac{1343}{12} \zeta_5 \Bigg)  \Bigg\} 
                               \nonumber\\
&+ 
                               {\dis{2 C_{F} n_{f} T_{F}}} \Bigg\{ - \frac{40}{3
                               \epsilon^3} 
                                + \frac{280}{9 \epsilon^2} + 
                               \Bigg( - \frac{1417}{27} +  2 \zeta_2
                               \Bigg) \frac{1}{\epsilon}  +
                               \frac{22021}{324} -  \frac{14}{3}
                               \zeta_2  -  \frac{82}{9} \zeta_3 
\nonumber\\
&+ 
                               \epsilon \Bigg( - \frac{238717}{3888} -
                               \frac{73}{12} \zeta_2 
                              + \frac{25}{12}
                               \zeta_2^2  + \frac{394}{27} \zeta_3 \Bigg) + 
                               \epsilon^2 \Bigg( -
                               \frac{290075}{46656}  +
                               \frac{6181}{144} \zeta_2 -
  \frac{499}{180} \zeta_2^2 
\nonumber\\
&- 
                               \frac{9751}{324} \zeta_3 + \frac{13}{6}
                               \zeta_2 \zeta_3  
                              - \frac{29}{6} \zeta_5 \Bigg) \Bigg\} 
                               + 
                               {\dis{C_{F}^2}} \Bigg\{ \frac{32}{\epsilon^4} -
                               \frac{48}{\epsilon^3} +  \Bigg( 62 -  8
                               \zeta_2 \Bigg)  \frac{1}{\epsilon^2} +
                               \Bigg( - \frac{113}{2}  
\nonumber\\
&+ \frac{128}{3}
                               \zeta_3 \Bigg) \frac{1}{\epsilon} +
                               \frac{581}{24} +  \frac{27}{2} \zeta_2 
                              - 
                               13 \zeta_2^2 - 58 \zeta_3 + 
                               \epsilon \Bigg( \frac{12275}{288} -
                               \frac{331}{24} \zeta_2  +
                               \frac{493}{30} \zeta_2^2  +
                               \frac{587}{6} \zeta_3  
\nonumber\\
&- 
                               \frac{56}{3} \zeta_2 \zeta_3 +
                               \frac{92}{5} \zeta_5 \Bigg) 
                               + 
                               \epsilon^2 \Bigg( - \frac{456779}{3456}
                               - \frac{2011}{96} \zeta_2 -
                               \frac{1279}{80} \zeta_2^2  + 
                               \frac{223}{20} \zeta_2^3 -
                               \frac{13363}{72} \zeta_3  
\nonumber\\
&- \frac{5}{2}
                               \zeta_2 \zeta_3  + 
                               \frac{652}{9} \zeta_3^2  
                              -
                               \frac{193}{30} \zeta_5  \Bigg)  \Bigg\} 
                               + 
                               {\dis{2 C_{A} n_{f} T_{F}}} \Bigg\{ - \frac{176}{9
                               \epsilon^2} +  \frac{1972}{27 \epsilon}
                               - \frac{1708}{9} +  \frac{80}{9} \zeta_2 + 
                               4 \zeta_3 
\nonumber\\
&+ 
                               \epsilon \Bigg( \frac{104858}{243}  -
                               \frac{853}{27} \zeta_2 
                               - \frac{2}{3}
                               \zeta_2^2  - \frac{1622}{27} \zeta_3
                               \Bigg) +  \epsilon^2 \Bigg( -
                               \frac{5369501}{5832} +  \frac{1447}{18}
                               \zeta_2 +  \frac{817}{45} \zeta_2^2 
\nonumber\\
&+ 
                               \frac{31499}{162} \zeta_3 + \frac{7}{3}
                               \zeta_2 \zeta_3  
                              + 19 \zeta_5 \Bigg) \Bigg\} 
                               + 
                               {\dis{C_{A} C_{F}}} \Bigg\{ \frac{220}{3 \epsilon^3}+ 
                               \Bigg( - \frac{1804}{9} + 4 \zeta_2
                               \Bigg) \frac{1}{\epsilon^2} +  \Bigg(
                               \frac{20777}{54} 
\nonumber\\
&-  19 \zeta_2 - 50
                               \zeta_3 \Bigg)  \frac{1}{\epsilon}  
                              -
                               \frac{397181}{648 } + \frac{161}{3}
                               \zeta_2 +  \frac{76}{5} \zeta_2^2 + 
                               \frac{1333}{9} \zeta_3 + \epsilon
                               \Bigg(  \frac{6604541}{7776} -
                               \frac{669}{8} \zeta_2  
\nonumber\\
&- 
                               \frac{5519}{120} \zeta_2^2 -
                               \frac{8398}{27} \zeta_3  
                              + \frac{89}{6}
                               \zeta_2 \zeta_3  - 
                               \frac{51}{2} \zeta_5 \Bigg) +
                               \epsilon^2 \Bigg(  -
                               \frac{93774821}{93312} +
                               \frac{20035}{288} \zeta_2 +  
                               \frac{33377}{360} \zeta_2^2 
\nonumber\\
&+
                               \frac{1793}{840} \zeta_2^3  +
                               \frac{390731}{648} \zeta_3 
                              -  
                               \frac{445}{12} \zeta_2 \zeta_3 -
                               \frac{425}{12} \zeta_3^2  +
                               \frac{641}{12} \zeta_5 \Bigg) \Bigg\}\,,
\end{align}

% **** FqJ1 ****

\begin{align}
  \label{eq:FqJ1}
  {\hat{\cal F}}^{J,(1)}_{q} &= {\dis{C_{F}}} \Bigg\{ - \frac{8}{\epsilon^2}
                               + \frac{6}{\epsilon}  - 2 + \zeta_2  +
                               \epsilon \Bigg( - 1  - \frac{3}{4}
                               \zeta_2  - \frac{7}{3} \zeta_3 \Bigg)  + 
                               \epsilon^2 \Bigg( \frac{5}{2} +
                               \frac{\zeta_2}{4}  + \frac{47}{80}
                               \zeta_2^2  + \frac{7}{4} \zeta_3 \Bigg) 
                               \nonumber\\
                             &+ 
                               \epsilon^3 \Bigg( - \frac{13}{4}  +
                               \frac{\zeta_2}{8}  - \frac{141}{320}
                               \zeta_2^2  - \frac{7}{12} \zeta_3 + 
                               \frac{7}{24} \zeta_2 \zeta_3  -
                               \frac{31}{20} \zeta_5  \Bigg) \Bigg\}\,,
\end{align}

% **** FqJ2 ****

\begin{align}
  \label{eq:FqJ2}
  {\hat{\cal F}}^{J,(2)}_{q} &={\dis{2 C_{F} n_{f} T_{F}}} \Bigg\{ - \frac{8}{3
                               \epsilon^3}  + \frac{56}{9 \epsilon^2}
                               + \Bigg( - \frac{47}{27} - \frac{2}{3}
                               \zeta_2 \Bigg)  \frac{1}{\epsilon} -
                               \frac{4105}{324}  + 
                               \frac{14}{9} \zeta_2  - \frac{26}{9}
                               \zeta_3  + \epsilon \Bigg(
                               \frac{142537}{3888}  
                               \nonumber\\
                             &- \frac{695}{108}
                               \zeta_2  + 
                               \frac{41}{60} \zeta_2^2 +
                               \frac{182}{27} \zeta_3 \Bigg)  + 
                               \epsilon^2 \Bigg( -
                               \frac{3256513}{46656}  +
                               \frac{21167}{1296} \zeta_2 
                               - \frac{287}{180} \zeta_2^2 - 
                               \frac{2555}{324} \zeta_3  
                               \nonumber\\
                             &-
                               \frac{13}{18} \zeta_2 \zeta_3  -
                               \frac{121}{30} \zeta_5  \Bigg) \Bigg\} 
                               + 
                               {\dis{C_{F}^2}} \Bigg\{ \frac{32}{\epsilon^4}
                               - \frac{48}{\epsilon^3}  + \Bigg( 34 -
                               8 \zeta_2 \Bigg)  \frac{1}{\epsilon^2}
                               + \Bigg( - \frac{5}{2} + \frac{128}{3}
                               \zeta_3 \Bigg)  \frac{1}{\epsilon} -
                               \frac{361}{8}  
                               \nonumber\\
                             &+ \frac{9}{2} \zeta_2 - 
                               13 \zeta_2^2 - 58 \zeta_3  + 
                               \epsilon \Bigg( \frac{3275}{32}  +
                               \frac{3}{8} \zeta_2  + \frac{171}{10}
                               \zeta_2^2  + \frac{503}{6} \zeta_3 - 
                               \frac{56}{3} \zeta_2 \zeta_3  +
                               \frac{92}{5} \zeta_5 \Bigg)  
                               \nonumber\\
                             &+ 
                               \epsilon^2 \Bigg( - \frac{20257}{128}
                               - \frac{793}{32} \zeta_2  -
                               \frac{2097}{80} \zeta_2^2  + 
                               \frac{223}{20} \zeta_2^3 -
                               \frac{4037}{24} \zeta_3  + \frac{27}{2}
                               \zeta_2 \zeta_3  + 
                               \frac{652}{9} \zeta_3^2 -
                               \frac{231}{10}  \zeta_5 \Bigg) \Bigg\} 
                               \nonumber\\
                             &+ 
                               {\dis{C_{A} C_{F}}} \Bigg\{ \frac{44}{3
                               \epsilon^3}  + 
                               \Bigg( - \frac{332}{9} + 4 \zeta_2
                               \Bigg) \frac{1}{\epsilon^2}  + \Bigg(
                               \frac{2545}{54}  + \frac{11}{3} \zeta_2
                               - 26 \zeta_3 \Bigg) \frac{1}{\epsilon}
                               - \frac{18037}{648} - \frac{47}{9}
                               \zeta_2  
                               \nonumber\\
                             &+ \frac{44}{5} \zeta_2^2 + 
                               \frac{467}{9} \zeta_3 + \epsilon \Bigg(
                               - \frac{221963}{7776}  -
                               \frac{263}{216} \zeta_2  -
                               \frac{1891}{120} \zeta_2^2  - 
                               \frac{2429}{27} \zeta_3 + \frac{89}{6}
                               \zeta_2 \zeta_3  - \frac{51}{2} \zeta_5
                               \Bigg)  
                               \nonumber\\
                             &+ 
                               \epsilon^2 \Bigg(
                               \frac{11956259}{93312}  +
                               \frac{38987}{2592} \zeta_2  +
                               \frac{9451}{360} \zeta_2^2  - 
                               \frac{809}{280} \zeta_2^3 +
                               \frac{92701}{648} \zeta_3  -
                               \frac{397}{36} \zeta_2 \zeta_3  - 
                               \frac{569}{12} \zeta_3^2  
                               \nonumber\\
                             &+ \frac{3491}{60} \zeta_5 \Bigg) \Bigg\}\,,
\end{align}

% **** FqJ3 ****

\begin{align}
  \label{eq:FqJ3}
  {\hat{\cal F}}^{J,(3)}_{q} &= {\dis{4 C_{F} n_{f}^2 T_{F}^{2}}} \Bigg\{ - \frac{128}{81
                               \epsilon^4}  + \frac{1504}{243
                               \epsilon^3}  + 
                               \Bigg( - \frac{16}{9}  - \frac{16}{9}
                               \zeta_2 \Bigg) \frac{1}{\epsilon^2}  + 
                               \Bigg( - \frac{73432}{2187}  +
                               \frac{188}{27} \zeta_2  
\nonumber\\
&-
                               \frac{272}{81} \zeta_3 \Bigg)
                               \frac{1}{\epsilon}  
                              +
                               \frac{881372}{6561}  - 26 \zeta_2  -
                               \frac{83}{135} \zeta_2^2   + \frac{3196}{243} \zeta_3 \Bigg\} 
                               + 
                               {\dis{C_{F}^3}} \Bigg\{ - \frac{256}{3
                               \epsilon^6}  + \frac{192}{\epsilon^5} +
                               \Bigg( - 208  
\nonumber\\
&+ 32 \zeta_2 \Bigg)
                               \frac{1}{\epsilon^4}  
                              + 
                               \Bigg( 88 + 24 \zeta_2  - \frac{800}{3}
                               \zeta_3 \Bigg)  \frac{1}{\epsilon^3}  +
                               \Bigg( 254  - 98 \zeta_2 +
                               \frac{426}{5} \zeta_2^2  + 552 \zeta_3
                               \Bigg) \frac{1}{\epsilon^2}  
\nonumber\\
&+ 
                               \Bigg( - \frac{5045}{6}  + 83 \zeta_2
                              - \frac{1461}{10} \zeta_2^2  -
                               \frac{2630}{3} \zeta_3  + 
                               \frac{428}{3} \zeta_2 \zeta_3  -
                               \frac{1288}{5} \zeta_5 \Bigg)
                               \frac{1}{\epsilon} + \frac{38119}{24}
                               + \frac{1885}{12} \zeta_2  
\nonumber\\
&+ 
                               \frac{8659}{40} \zeta_2^2  
                              -
                               \frac{9095}{252} \zeta_2^3   + 1153
                               \zeta_3 - 35 \zeta_2 \zeta_3  - 
                               \frac{1826}{3} \zeta_3^2  - \frac{562}{5} \zeta_5 \Bigg\} 
                               + 
                               {\dis{2 C_{F}^2 n_{f} T_{F}}} \Bigg\{ \frac{64}{3
                               \epsilon^5}  - \frac{592}{9 \epsilon^4}
                               \nonumber\\
&+ \Bigg( \frac{1480}{27}  
                              + \frac{8}{3}
                               \zeta_2 \Bigg)  \frac{1}{\epsilon^3} + 
                               \Bigg( \frac{7772}{81} - \frac{266}{9}
                               \zeta_2  + \frac{584}{9} \zeta_3 \Bigg)
                               \frac{1}{\epsilon^2} + 
                               \Bigg( - \frac{116735}{243}  +
                               \frac{2633}{27} \zeta_2  
\nonumber\\
&-
                               \frac{337}{18} \zeta_2^2  
                              -
                               \frac{5114}{27} \zeta_3 \Bigg)
                               \frac{1}{\epsilon}  +
                               \frac{3396143}{2916}  -
                               \frac{32329}{162} \zeta_2  + 
                               \frac{8149}{216} \zeta_2^2 + 
                               \frac{39773}{81} \zeta_3  -
                               \frac{343}{9} \zeta_2 \zeta_3  
\nonumber\\
&+ \frac{278}{45} \zeta_5 \Bigg\} 
                              + 
                               {\dis{C_{A}^2 C_{F}}} \Bigg\{ - \frac{3872}{81
                               \epsilon^4}  + \Bigg( \frac{52168}{243}
                               - \frac{704}{27} \zeta_2 \Bigg)
                               \frac{1}{\epsilon^3}  + 
                               \Bigg( - \frac{117596}{243} -
                               \frac{2212}{81} \zeta_2  
\nonumber\\
&-
                               \frac{352}{45} \zeta_2^2  
                              +
                               \frac{6688}{27} \zeta_3 \Bigg)
                               \frac{1}{\epsilon^2}  + \Bigg(
                               \frac{1322900}{2187}  +
                               \frac{39985}{243} \zeta_2  - 
                               \frac{1604}{15} \zeta_2^2 -
                               \frac{24212}{27} \zeta_3  +
                               \frac{176}{9} \zeta_2 \zeta_3  
\nonumber\\
&+ 
                               \frac{272}{3} \zeta_5 \Bigg)
                               \frac{1}{\epsilon}  
                              +
                               \frac{1213171}{13122} -
                               \frac{198133}{729} \zeta_2  +
                               \frac{146443}{540} \zeta_2^2  - \frac{6152}{189} \zeta_2^3 + 
                               \frac{970249}{486} \zeta_3 -
                               \frac{926}{9} \zeta_2 \zeta_3  
\nonumber\\
&-
                               \frac{1136}{9} \zeta_3^2  
                              + \frac{772}{9} \zeta_5  \Bigg\} 
                               + 
                               {\dis{2 C_{A} C_{F} n_{f} T_{F}}} \Bigg\{
                               \frac{1408}{81 \epsilon^4}  + \Bigg( -
                               \frac{18032}{243}  + \frac{128}{27}
                               \zeta_2 \Bigg)  \frac{1}{\epsilon^3} + 
                               \Bigg( \frac{24620}{243} 
\nonumber\\
&+
                               \frac{1264}{81} \zeta_2  
                              -
                               \frac{1024}{27} \zeta_3 \Bigg)
                               \frac{1}{\epsilon^2}  + 
                               \Bigg( \frac{212078}{2187} -
                               \frac{16870}{243} \zeta_2  +
                               \frac{88}{5} \zeta_2^2  +
                               \frac{12872}{81} \zeta_3 \Bigg)
                               \frac{1}{\epsilon}  -
                               \frac{5807647}{6561}  
                               \nonumber\\
                             &+
                               \frac{299915}{1458} \zeta_2  - 
                               \frac{5492}{135} \zeta_2^2 - 
                               \frac{42941}{81} \zeta_3 +
                               \frac{422}{9} \zeta_2 \zeta_3  - \frac{28}{3} \zeta_5 \Bigg\} 
                               + 
                               {\dis{C_{A} C_{F}^2}} \Bigg\{ - \frac{352}{3 \epsilon^5} + 
                               \Bigg( \frac{3448}{9}  
                               \nonumber\\
                             &- 32 \zeta_2
                               \Bigg) \frac{1}{\epsilon^4}  + \Bigg( -
                               \frac{16948}{27}  + \frac{28}{3}
                               \zeta_2 + 208 \zeta_3 \Bigg)
                               \frac{1}{\epsilon^3}  + \Bigg(
                               \frac{44542}{81}  + \frac{1127}{9}
                               \zeta_2  - \frac{332}{5} \zeta_2^2 
                               \nonumber\\
                             &- 
                               840 \zeta_3 \Bigg) \frac{1}{\epsilon^2}
                               + \Bigg( \frac{149299}{486}  -
                               \frac{12757}{54} \zeta_2  +
                               \frac{9839}{36} \zeta_2^2  + 
                               \frac{5467}{3} \zeta_3 - \frac{430}{3}
                               \zeta_2 \zeta_3  + 284 \zeta_5 \Bigg)
                               \frac{1}{\epsilon}  
                               \nonumber\\
                             &-
                               \frac{15477463}{5832}  +
                               \frac{21455}{324} \zeta_2  -
                               \frac{1002379}{2160} \zeta_2^2  - 
                               \frac{18619}{1260} \zeta_2^3 -
                               \frac{51781}{18} \zeta_3  +
                               \frac{910}{9} \zeta_2 \zeta_3  + 
                               \frac{1616}{3} \zeta_3^2 
                               \nonumber\\
                             &- 
                               \frac{3394}{45} \zeta_5 \Bigg\}\,. 
\end{align}

\bibliography{pScalar} \bibliographystyle{utphysM}
  
\end{document}